\documentclass[]{IEEEtran}
\usepackage{amsmath}
\usepackage{amsthm}
\usepackage{amsfonts}
\usepackage{amssymb}
\usepackage{makeidx}
\usepackage{cite}
\usepackage{graphicx,bigints}
\usepackage{mathtools}
\usepackage{cases}
\usepackage{color,soul}
\usepackage{hyperref}
\usepackage{bbm}
\usepackage[normalem]{ulem}
\usepackage{braket}
\usepackage{textcomp,gensymb}
\usepackage{enumitem}
\usepackage{epstopdf}
\usepackage{xcolor}
\usepackage{lipsum}

\usepackage{multirow}

\usepackage[makeroom]{cancel}

\makeatletter
\DeclareFontFamily{U}{tipa}{}
\DeclareFontShape{U}{tipa}{m}{n}{<->tipa10}{}
\newcommand{\arc@char}{{\usefont{U}{tipa}{m}{n}\symbol{62}}}%

\newcommand{\arc}[1]{\mathpalette\arc@arc{#1}}

\newcommand{\arc@arc}[2]{%
	\sbox0{$\m@th#1#2$}%
	\vbox{
		\hbox{\resizebox{\wd0}{\height}{\arc@char}}
		\nointerlineskip
		\box0
	}%
}
\makeatother

\newtheorem{theorem}{Theorem}
\newtheorem{corollary}{Corollary}
\newtheorem{lemma}{Lemma}

\newtheorem{example}{Example}

\newtheorem{remark}{Remark}

\DeclareMathOperator{\cA}{\mathcal{A}}

\DeclareMathOperator{\cT}{\mathcal{T}}

\DeclareMathOperator{\cO}{\mathcal{O}}

\DeclareMathOperator{\cS}{\mathcal{S}}
\DeclareMathOperator{\cF}{\mathcal{F}}

\DeclareMathOperator{\cR}{\mathcal{R}}

\DeclareMathOperator{\SNR}{\textrm{SNR}}

\DeclareMathOperator{\bP}{\mathbf{P}}
\DeclareMathOperator{\ind}{\mathbbm{1}}
\DeclareMathOperator{\bE}{\mathbf{E}}

\newcommand*\diff{\mathop{}\!\mathrm{d}}

\newcommand*\nnb{\nonumber}

\newcommand{\overbar}[1]{\mkern 1.5mu\overline{\mkern-1.5mu#1\mkern-1.5mu}\mkern 1.5mu}

\definecolor{sandy}{HTML}{E6E2AF}
\definecolor{stone}{HTML}{A7A37E}
\definecolor{beach}{HTML}{EFECCA}
\definecolor{ocean}{HTML}{046380}
\definecolor{diver}{HTML}{002F2F}

\definecolor{Firenze1}{HTML}{468966}
\definecolor{Firenze2}{HTML}{FFF0A5}
\definecolor{Firenze3}{HTML}{FFB03B}
\definecolor{Firenze4}{HTML}{B64926}
\definecolor{Firenze5}{HTML}{8E2800}
\definecolor{mediumpersianblue}{rgb}{0.0, 0.4, 0.65}
\definecolor{hongik}{HTML}{004498}
\definecolor{cobalt}{rgb}{0.0, 0.28, 0.67}
\definecolor{burntorange}{rgb}{0.8, 0.33, 0.0}

\definecolor{ultramarineblue}{rgb}{0.25, 0.4, 0.96}

\title{Analysis of a Delay-Tolerant Data Harvest Architecture Leveraging Low Earth Orbit Satellite Networks}

\author{Chang-Sik Choi~\IEEEmembership{Member IEEE} 
	\IEEEcompsocitemizethanks{\IEEEcompsocthanksitem{Chang-Sik Choi is an Assistant Professor of Dept. of EE, Hongik University, South Korea. (chang-sik.choi@hongik.ac.kr)
		} }
}
\begin{document}
	\maketitle 
	
	\begin{abstract}
Reaching all regions of Earth, low Earth orbit (LEO) satellites can harvest delay-tolerant data from remotely located users on Earth without ground infrastructure. This work aims to assess a data harvest network architecture where users generate data and LEO satellites harvest data from users when passing by. By developing a novel stochastic geometry Cox point process model that simultaneously generates orbits and the motion of LEO satellite harvesters on them, we analyze key performance indices of such a network by deriving the following: (i) the average fraction of time that the typical user is served by LEO satellite harvesters, (ii) the average amount of data uploaded per each satellite pass, (iii) the maximum harvesting capacity of the proposed network model, and (iv) the delay distribution in the proposed network. These key metrics are given as functions of key network variables such as $\lambda$ the mean number of orbits and $\mu$ the mean number of satellites per orbit. Providing rich comprehensive analytical results and practical interpretations of these results, this work assesses the potential of the delay-tolerant use of LEO satellites and also serves as a versatile framework to analyze, design, and optimize delay-tolerant LEO satellite networks.
\end{abstract}


	\section{Introduction}\label{S:1}

\subsection{Motivation}

\IEEEPARstart{L}{EO} satellites have the potential to provide connectivity to millions of devices on Earth without the need for nearby ground infrastructure \cite{Iridium,6934544,8700141,FCCKuiper,FCCBoeing}. Unlike GEO satellites, which remain stationary in their geostationary orbits, LEO satellites rotate around Earth along their orbits, providing ubiquitous coverage to users worldwide and establishing connections between the LEO satellites and those users \cite{8002583,9711564,9755278,9861699,9970355,10209551}.

Among the wide range of services that LEO satellite networks can provide, one of the most critical applications is remote data sensing or data harvesting \cite{8002583,9711564,10108903}. This involves various devices such as IoT devices, disconnected mobiles, or UAVs that generate delay-tolerant data while they may be far from ground infrastructure. When these devices are within the range of LEO satellites, they can upload their data to the best available satellite data harvester. Efficient inter-satellite and satellite-to-ground gateway links are utilized to deliver the data to their final destinations or the Internet.

However, small devices such as IoT devices or disconnected mobile handsets may experience additional delays due to their limited power, finite communication range, or sparse LEO satellite distribution. In some areas, the density of the LEO satellite constellation may be low, making data harvesting from such devices very challenging. To build a successful data harvesting architecture leveraging LEO satellite harvesters, it is crucial to understand the data harvesting capacity and the delay distribution which of them are jointly affected by key distributional factors such as the number of satellites, the number of orbital planes, and the speeds of LEO satellites on their orbits. This paper aims to model and analyze a data harvesting architecture based on LEO satellites to assess the potential of such a network and provide design insights and intuitions.

\subsection{Related Work}

Providing wide coverage areas without ground infrastructure, non-terrestrial satellite networks have received a lot of attention from industries and academia \cite{5959951}. To name a few, \cite{6174409,7060478} discusses the coverage-related questions of LEO satellite networks; \cite{8352859} considered the routing problems in satellite networks; \cite{8473417,8473415,8626457,8700142} discussed the integration of satellite networks and 5G infrastructure; \cite{7572177} analyzed the inter-satellite links and information routing problems; \cite{8571192} investigated uplink communications or data harvesting from ground data devices to satellites. In particular, since LEO satellites are capable of covering everywhere on Earth, LEO satellite networks are one of the best candidates for retrieving data from devices all around Earth, especially when they are far from any ground infrastructure.

In the context of data harvesting based on LEO satellites, achieving reasonable delays and reliable communications is possible when there are many LEO satellites within the communication ranges of users \cite{9526866}. However, in practice, factors such as limited transmit powers, short communication distances, sparse orbital planes, or low numbers of LEO satellites may lead to additional delays for users to find their LEO satellites and establish uplink communications \cite{9755278,9970355}. To address this practical challenge in delay-tolerant LEO satellite networks, various technologies such as beam sweeping and efficient resource management techniques have been developed \cite{9526866}. Similarly, this work aims to study delay-tolerant data harvesting architectures leveraging LEO satellite harvesters by focusing on the influence of distributional aspects on the network's large-scale performance behavior.

To analyze the impact of geometric variables in LEO satellite networks, \cite{9079921,9177073,9218989,9497773,9678973,9841569,9838263,9861782} have employed stochastic geometry models \cite{daley2007introduction,chiu2013stochastic,baccelli2010stochastic}, which have been successfully used in the analysis of cellular or vehicular networks \cite{5226957}. These stochastic geometry models were used to describe the locations of network components and help one to obtain typical network performance as a function of network parameters. By providing typical performance as network parameters and eliminating the need for complex and time-consuming system-level simulations, these models have been well-received in the literature on the analysis of LEO satellite networks. Previous studies \cite{9079921,9177073,9218989,9497773} have used binomial point processes to model the locations of LEO satellites as uniformly and randomly distributed points on a sphere. For instance, \cite{9678973} analyzed the coverage probability of downlink communications from LEO satellites to ground users by ignoring interference. On the other hand, \cite{9838263,9861782} considered inter-satellite downlink interference and analyzed downlink communications under Nakagami fading. Uplink communications were investigated in \cite{9509510}. In these works \cite{9079921,9177073,9218989,9497773,9678973,9841569,9838263,9861782}, a snapshot of the network geometry was used to replicate the distribution of the LEO satellite constellation at any given time.

However, those static models \cite{9079921,9177073,9218989,9497773,9678973,9841569,9838263,9861782} are not suitable for analyzing delay-tolerant data harvesting architectures as they do not incorporate the motion of LEO satellites on their orbital planes. We need an analytical framework handling the dynamic nature of a delay-tolerant LEO satellite network while revealing the network's time-domain characteristics. Recently, \cite{choi2022analytical,choi2023Cox,choi2023Cox2} developed Poisson orbit Cox point processes that jointly model orbital planes and LEO satellites exclusively on these orbital planes. Unlike the previous modeling approach based on binomial or Poisson point processes, these Cox point process models not only introduce randomness to the locations of LEO satellites but also generate their orbital trajectories. While \cite{choi2022analytical} studied downlink communications from LEO satellites to ground users and \cite{choi2023Cox} provided important statistical properties and attributes, the full potential of the Cox point process, which allows for investigating the motion and dynamics of LEO satellites by generating their trajectories, has not been fully utilized in these papers. In the present work, we adapt the Cox point process modeling approach and exploit its full potential to quantify the time-domain behavior of LEO satellite networks, leveraging the fact that LEO satellites move on the orbital planes produced by the Cox satellite point process. To the best of the authors' knowledge, the delay-tolerant nature of LEO satellite networks has not been evaluated using stochastic geometry analysis yet due to the lack of an appropriate model. Therefore, within the context of stochastic geometry, this work is the first attempt to analyze the harvesting capacity and association delay of the LEO satellite network architecture while emphasizing the distribution of LEO satellites and their motion on orbital planes.

\subsection{Theoretical Contributions}
The theoretical contributions of this paper are as follows:
\begin{itemize}
\item \emph{Modeling LEO satellites and their motion}: In this work, we employ a Cox point process to integrate the distribution and dynamics of LEO satellites on their orbits. In previous work  \cite{choi2022analytical,choi2023Cox,choi2023Cox2}, the distribution of LEO satellites on orbital planes was fully characterized. Nevertheless, the orbital planes produced by the Cox point process were not fully utilized to describe the dynamics of LEO satellites on their orbits over time. We leverage the conditional structure of the Cox point process so that the LEO satellites on every orbit are forced to follow their orbits with a constant speed $ v $. This work identifies the existing spatial correlation of satellites on their orbits to characterize their motion so that the time domain behavior of a LEO satellite network can be analyzed in a single all-embracing framework. 

\item \emph{Analysis of harvest time fraction, harvest amount of data per pass, harvesting capacity, and user delay distribution}: Leveraging the random distribution of LEO satellites and their correlated movements on orbits, we first analyze the average fraction of time that the typical user can communicate with LEO satellite data harvesters. Then, assuming Nakagami fading for uplink communication from users to satellite harvesters, we derive the amount of uploaded data from a typical network user to its LEO satellite during the LEO satellite's passage. This metric is useful as it provides the expected amount of uploaded data from each network user per satellite pass, particularly when the number of LEO satellites is low. To examine the harvesting capability of LEO satellite networks in more dense LEO satellite scenarios, we evaluate the harvesting capacity, which is defined as the achievable rate of typical uplink communication. This capacity corresponds to the long-term maximum data rate (bps/sec) processed and handled by the LEO satellite harvesters on orbits. Next, by identifying the dominant factors that determine the delay or wait time for users, we derive the delay distribution of the typical user in the network. 

\item \emph{Providing key insights and design principles}: Based on analytical and experimental results, this paper provides practical insights on how to design and optimize delay-tolerant data harvesting architectures using LEO satellites.
First, the analysis of the harvest time fraction reveals that the temporal availability of LEO satellites is heavily influenced by the number of orbital planes. This insight suggests that a deployment scenario with a greater number of orbital planes, even with the same number of LEO satellites, may achieve better overall energy efficiency compared to a scenario with fewer orbital planes.
Additionally, we express the harvesting capacity as a function of the number of orbital planes and the number of satellites. This measure directly reflects the quality of service that LEO satellite network operators can offer to ground users. By understanding this relationship, network operators can make informed decisions to optimize their services.
From the delay analysis, we explicitly outline the advantages of having more or fewer orbital planes. This information can be utilized by network operators to adjust or improve the user delay distribution, thereby enhancing the overall performance of the proposed architecture.

\end{itemize}

\section{System Model}
\subsection{Spatial Distribution}\label{S:2-A}
Without loss of generality, Earth center is at the origin $(0,0,0)$ and its radius is denoted by $r_e$; the reference plane is the $xy$-plane; the longitudinal reference direction is the $x$-axis. Every orbit is a circle of radius $r_o$ (i.e., satellite altitude $r_a = r_o-r_e$) and it's center is the origin. An orbit is completely characterized by its longitude and inclination. Specifically, the longitude is the angle from the $x$-axis to the direction of the ascending node, measured on the $xy$-plane. The inclination is the angle between the orbital plane and the $xy$-plane.  The argument of a satellite is defined as the angle between the ascending node and the satellite, measured on the orbital plane. 

\begin{figure}
	\centering
 	\includegraphics[width=.65\linewidth]{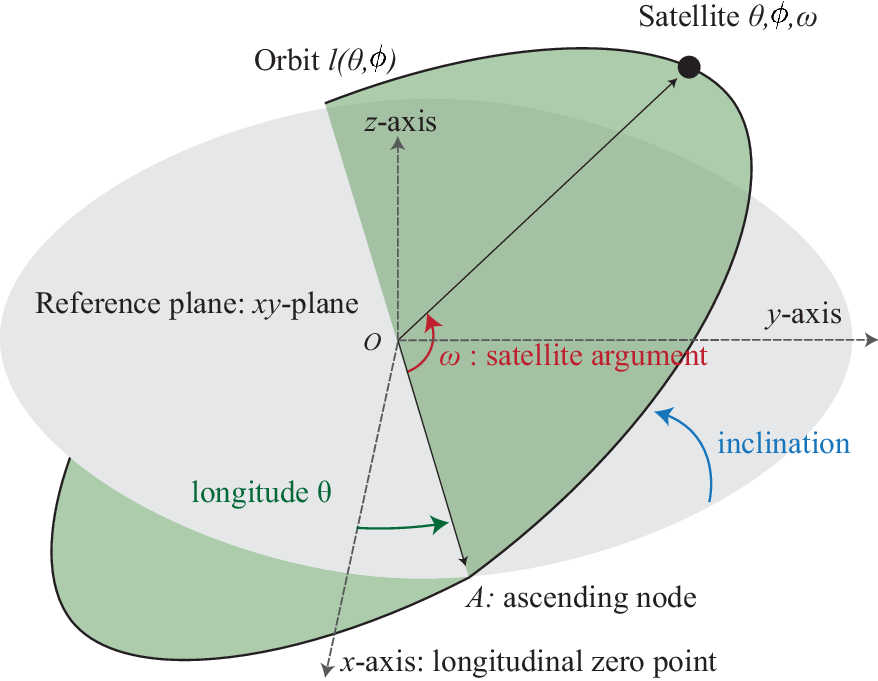}
	\caption{longitude $\theta$, inclination $\phi$, and satellite's argument angle $\omega$.}
	\label{fig:defineangles}
\end{figure}

In this work, the orbits of the LEO satellite networks are modeled as a Poisson point process $\cO $ of intensity $\frac{\lambda }{2\pi}\sin(\phi)$ on a rectangle $\cR = [0,\pi]\times[0,\pi],$ where the first coordinate corresponds to the orbit's longitude $\theta$ and the second coordinate corresponds to the orbit's inclination $\phi$.  For instance, a point $(\theta,\phi)=(\pi/2, \pi/3)$ denotes an orbit whose longitude is $\pi/2$ and inclination is $\pi/3$. The average number of orbits is $\lambda.$ We write 
\begin{equation}
	\cO = \sum_{i}\delta_{\theta_{i},\phi_{i}},
\end{equation}
where $\delta_{(\cdot,\cdot)}$ is a Dirac measure. An orbit with angle $(\theta,\phi)$ is denoted by $l(\theta,\phi).$ See Fig. \ref{fig:defineangles} for the angles defined in this paper. 

\begin{figure}
	\centering
		\includegraphics[width=1\linewidth]{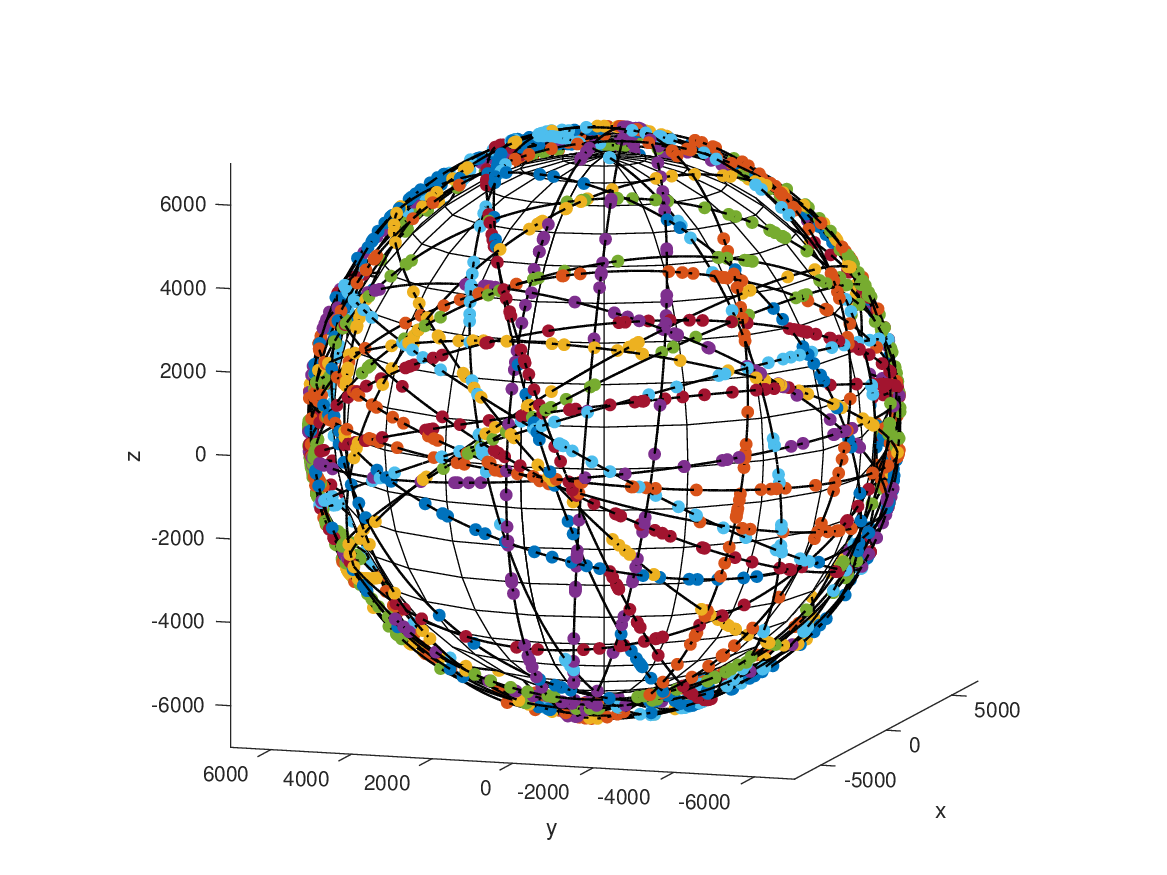}
	\caption{Illustration of the proposed network with $\lambda=40$ and $\mu=70$. There are $2800$ satellites on average.}
	\label{fig:deploy2515}
\end{figure}

\begin{figure}
	\centering
	\includegraphics[width=1\linewidth]{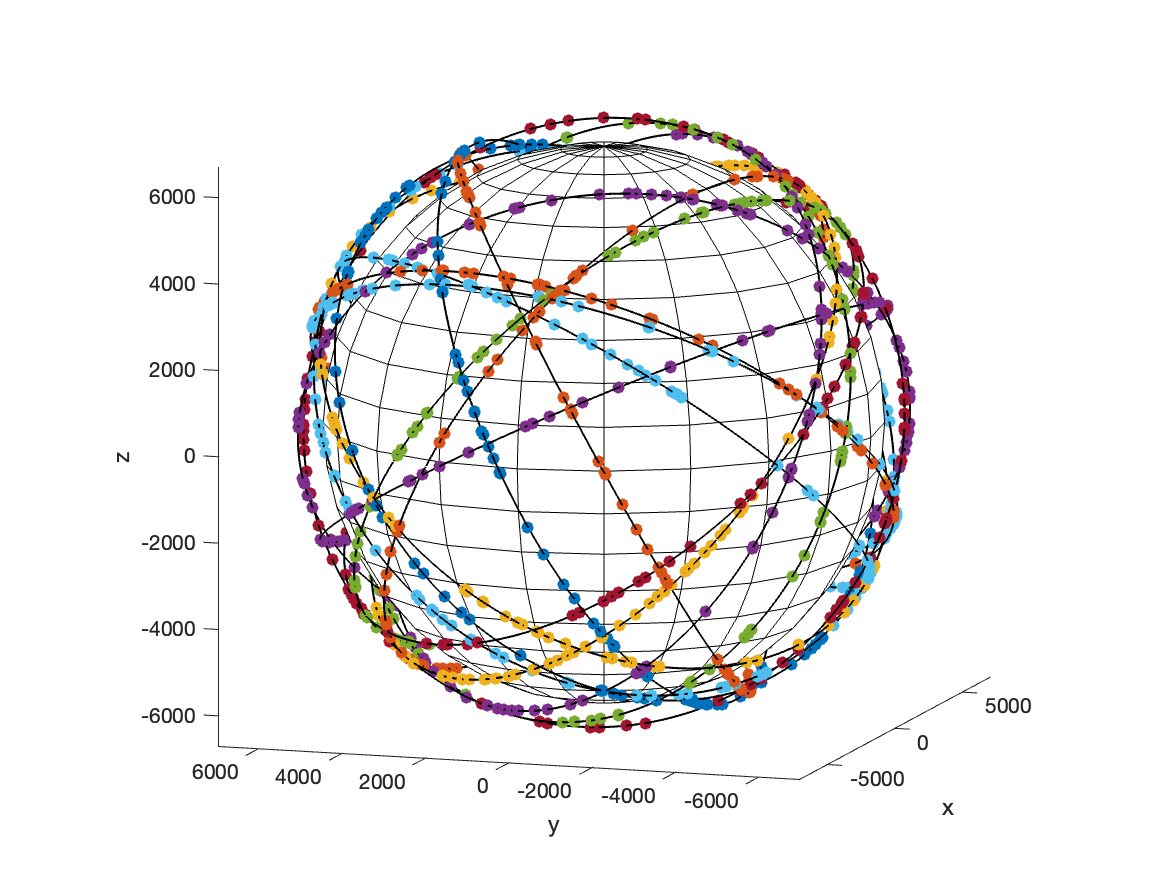}
	\caption{Illustration of the proposed network with $\lambda=15$ and $\mu=80$. There are $1200$ satellites on average.}
	\label{fig:deploy5030}
\end{figure}

Conditionally on the orbit process $\cO$,  LEO satellites on each orbit ${(\theta_{i},\phi_{i})}$ are modeled as a Poisson point process $\psi_{\theta_{i},\phi_{i}} $ of intensity $\mu/(2\pi r_o)$ on each orbit. In other words, on each orbit, there are $\mu$ number of LEO satellites on average. Since the locations of LEO satellites follow the Poisson point process, conditionally on the number of LEO satellites per orbit, the arguments $ \omega $ of LEO satellites are uniformly distributed over $0$ and $2\pi.$ The argument angle $\omega$ is the angle from the ascending point to the satellite, measured on its orbit. Collectively, the locations of satellites are given by a Cox point process $\Psi$ defined as follows: 
\begin{equation}
	\Psi = \sum_{i} \psi_{\theta_{i},\phi_{i}}.  
\end{equation}
We assume that LEO satellites on each orbit move along the orbit that they are located at the speed of $v_s$. Figs. \ref{fig:deploy2515} -- \ref{fig:deploy5030} illustrate the proposed satellite Cox point process with various $\lambda$ and $\mu$. \cite{choi2022analytical,choi2023Cox} proved that the Cox  point process constructed by a Poisson point process of density $\lambda\sin(\phi)/2$ on the rectangle $\cR=[0,\pi]\times[0,\pi]$ is rotation invariant, namely its probability distribution is preserved by the rotation of the reference plane. \footnote{Similar to this approach, a Cox point process was employed to model vehicles on roads in \cite{8357962,8419219}, effectively capturing the geometric constraint that vehicles are exclusively on roads.} 
%

To model data-generating users far from ground infrastructure, we assume that network users are randomly distributed on the surface of Earth according to a Poisson point process \cite{daley2007introduction,chiu2013stochastic,baccelli2010stochastic}. To incorporate the geographic characteristic of such users, those users are assumed to be independent of the locations of the satellites or orbits.

\subsection{Uplink Communication Range Gamma}\label{S:II-B}
As mentioned in Section \ref{S:1}, the power-limited or remotely-located users can upload their data to LEO satellites as long as they are within certain distances \cite{8626457,8700141,9970355}. This limitation arises because of various following factors: (i) transmit power constraints at users, (ii) minimum energy detection threshold at LEO satellites, and (iii) dynamic antenna array configuration of LEO satellites. To differentiate these factors with distributional parameters---such as $\lambda$ and $\mu$, this paper combines these factors into a single variable $\gamma$, which represents the maximum uplink communication distance between ground users and LEO satellites.

\begin{figure}
	\centering
	\includegraphics[width=0.5\linewidth]{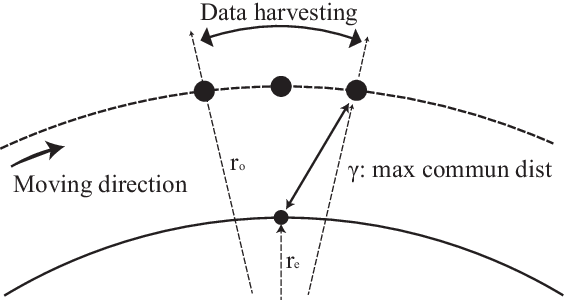}
	\caption{$\gamma$ is the maximum distance from the ground device to LEO satellite within which the data harvesting occurs. }
	\label{fig:gammaremark}
\end{figure}

In other words, a ground user can communicate with a single LEO satellite only when the distance from it to the LEO satellite harvester is less than $\gamma.$ Fig. \ref{fig:gammaremark} illustrates the area where the LEO satellite can collect data from the user. Due to the altitude of the satellite harvesters and the curvature of their trajectories, the value of $\gamma$ lies within the range of $[r_a, \sqrt{r_o^2-r_e^2}]$.

Besides, from the typical user at $(0,0,r_e), $ all the LEO satellite at distances less than $\gamma$ from the typical user must have their azimuth angles less than $\xi$. We use the Cosine law to get 
\begin{equation}
\xi = \arccos\left(\frac{r_e^2+r_o^2-\gamma^2}{2 r_e r_o}\right). \label{azi_sensing}
\end{equation}

\subsection{Propagation Model}
The received signal power of the uplink communication from the typical user to the LEO satellite at distance $ d$ (unit: meter and $d\geq 1$) is modeled as 
\begin{equation}
	p g H d^{-\alpha}, \label{PL}
\end{equation}
where $p$ is the transmit signal power, $g$ is an aggregate antenna gain based on the transmit and receive antenna arrays, $\alpha$ is the path loss exponent greater or equal to $2$, and $H$ is the random variable accounting for small-scale fading. To emphasize the interplay of network parameters and to ease the analysis, this paper considers $p,g,\alpha$ as fixed constants. For small-scale fading, this paper considers Nakagami-$m$ fading \cite{9861782}, which has been used in the analysis of LEO satellite communications\cite{9838263,9861782}. The distribution functions of the random variables $\sqrt{H}$ and $H$ are 
	 		\begin{align}
	f_{\sqrt{H}}(x;\,m,\Omega=1 )&={\frac {2m^{m}}{\Gamma (m)}}x^{2m-1}\exp \left(-{{m}}x^{2}\right) &\forall x\geq 0,\nnb\\	 			
	\bP(H>x) &= 
	e^{-mx}\sum\limits_{k=0}^{m-1} \frac{(m x)^k}{k!} &\forall x\geq 0,
	\label{CDF of H}
\end{align}
respectively, where $ m\geq 1 $, $ \Gamma(\cdot) $ is the Gamma function. With $ m=1$, Nakagami-$ m $ fading is Rayleigh fading and $ H $ follows an exponential random variable with mean $ 1 $.

\begin{table}
	\centering
	\caption{Network Variables}
	\begin{tabular}{|c|c|}		
		\hline
		Variables & Description \\ \hline 
		$\lambda$ & Average number of orbits \\ \hline
		$\mu$ & Average number of harvesters per orbit \\ \hline		
			$r_e$ & Earth radius \\ \hline
		$r_o$ & Radii of orbits \\ \hline
		$r_a$ & Orbit's altitude \\ \hline  
		$\cO $ & Orbit process  \\\hline  		
		$ l(\theta,\phi) $ & An orbit with angles $\theta$ and $\varphi$\\\hline
		$\psi_{\theta,\phi}$ & Satellite harvesters on the orbit $ l(\theta,\phi) $\\\hline 
		$\Psi$ & All satellite harvesters \\\hline 
		$\gamma$ & Communication range \\\hline 
		$v_s$ & Satellite speed ($\omega_s: $ angular speed)  \\\hline 
		$\cS_{\gamma}$ & Spherical cap with distance less than $\gamma$\\\hline 
				$\cA(\theta,\phi)$ & Intersection of $\cS_{\gamma}$ and $l(\theta,\varphi)$\\\hline 
				$p$ & Transmit power \\\hline
				$g$ & Aggregate antenna gain \\ \hline
					$d$ & Link distance \\ \hline
					$\alpha$ & Path loss exponent \\ \hline
					$\tau$ & SNR threshold \\\hline 
					$\sigma^2$ & Noise power \\\hline
					$B_w$ & Bandwidth \\\hline 
	\end{tabular}
\end{table}

\subsection{Performance Metric}
We investigate various practical time-domain metrics that capture the performance of the proposed delay-tolerant LEO satellite network. The following analysis is possible since our model integrates the locations of satellites and their dynamics, creating a unifying framework. 
\begin{itemize}
	\item \emph{Harvest time fraction}: This metric measures the amount of fractional time that the typical user is able to see any of LEO satellites, by quantifying the event that the typical user has at least one LEO satellite within distance $\gamma.$ 
	A higher $\gamma$ value indicates more LEO satellites within the communication range, resulting in a higher harvest time fraction. Conversely, with a smaller $\gamma$, the harvest time fraction approaches zero. The network's distributional parameters, such as $\lambda$ and $\mu$, also affect the harvest time fraction.
	
	\item \emph{Harvest data per pass and harvesting capacity}: Due to the movement of LEO satellites on their orbits, the distance between users and LEO satellites varies over time. To assess the data harvesting capability of this dynamic delay-tolerant architecture, this paper examines two performance metrics: (i) harvest data per pass and (ii) harvesting capacity. Both metrics account for the temporal variation of the links between users and associated LEO satellites, as well as the corresponding signal attenuation.
	\item \emph{Delay distribution}: Since LEO satellites are randomly distributed on orbits and in motion, users may experience waiting time to find suitable LEO satellites. Leveraging the constant speed of LEO satellites, this paper characterizes the delay as the time the typical user needs to wait until it finds at least one LEO satellite within distance $\gamma$.

\end{itemize}

\subsection{Preliminary results}\label{S:2-Prelim}
\begin{itemize}
	\item \emph{The length of an orbit on a spherical cap}: given the typical user is located at $(0,0,r_e)$, the spherical cap $\cS_\gamma $ is defined as $\cS_\gamma = \{(x,y,z)\in\cS| \|(x,y,z)-(0,0,r_e)\| \leq \gamma \}$  where $\cS$ is a sphere of radius $r_o$ centered at the origin and $(0,0,r_e)$ is the location of the typical user. Since $\gamma$ is the maximum uplink communication distance, the spherical cap $\cS_\gamma$ is the subset of a sphere containing the LEO satellites that could harvest data from the typical user. Fig. \ref{fig:figure3v2} illustrates the spherical caps $ \cS_{\gamma}$. 
\begin{figure}
	\centering
	\includegraphics[width=0.7\linewidth]{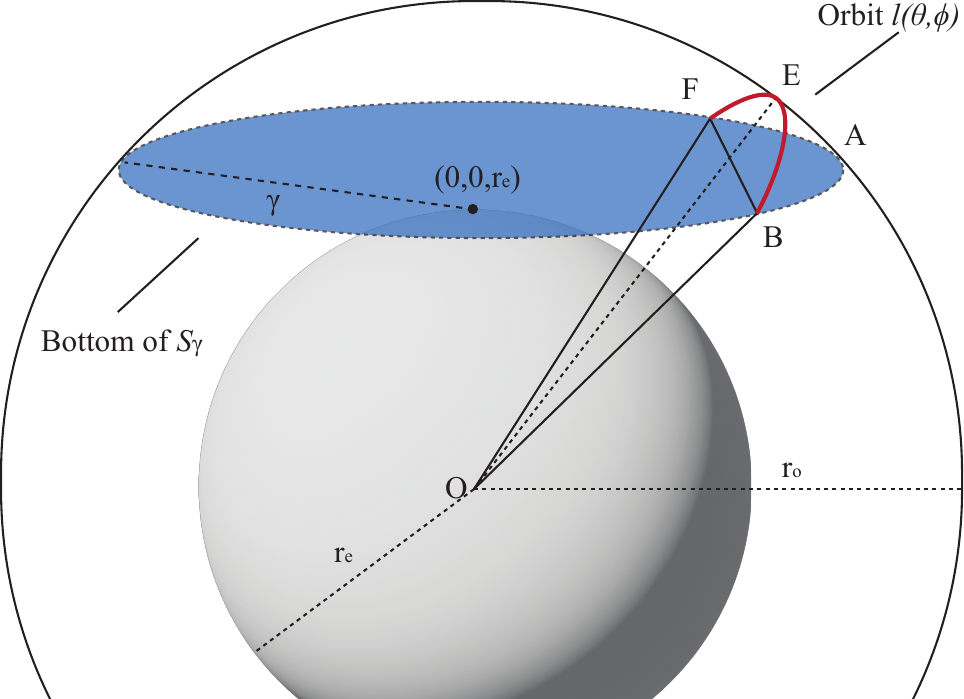}
	\caption{Illustration of the spherical cap $\cS_\gamma$. In the figure, $\cA(\theta,\phi)$ is  given by the red arc $  BF$. }
	\label{fig:figure3v2}
\end{figure}
 By the definition, the orbits with angles $\pi/2-\xi<\phi<\pi/2+\xi$  will meet the cap $ \cS_\gamma$.  Let $\cA(\theta,\phi)$ be the segment of orbit $l(\theta,\phi)$ intersecting the cap $\cS_\gamma.$ Then, its length is given by \begin{equation}
 	\|\cA(\theta,\phi)\|=2 r_o \arcsin \left(\sqrt{1-\cos^2(\xi)\csc^2(\phi)}\right),\label{arc:length}
 \end{equation}
$ $ where $\xi$ is Eq. \eqref{azi_sensing}. See \cite{choi2022analytical} for proof. 
\item \emph{Distance to a satellite}: Based on \cite{choi2022analytical}, the distance from the typical user at $(0,0,r_e)$ to a satellite parameterized by angles $(\theta,\phi,\omega)$ is given by 
\begin{equation}
	\sqrt{r_o^2-2r_er_o\sin(\omega)\sin(\phi)+r_e^2}.
\end{equation}
\end{itemize}

\section{Number of Harvesters and Network Geometry}
\subsection{Cox Property}
\begin{lemma}\label{P1}
	The proposed satellite Cox point process is time invariant. In other words, the distributions of $\Psi$ at any given times are identical. 
\end{lemma}
\begin{IEEEproof}
	After time $t>0,$ the longitude of every orbit of $\cO$ is rotated by $\omega_e t$  where $\omega_e$ is the angular speed of Earth's rotation. Since the proposed orbit process $\cO$ is invariant w.r.t. the longitudinal rotation of the reference plane, we conclude that the orbit process of any given time has the same distribution as the orbit process at time $0.$ Therefore, the orbit process is time invariant.  
	
	The satellite point process $\Psi$ is a Cox point process. Therefore, conditionally on $\cO,$ the satellite point process on each orbit follows a Poisson point process of intensity $\mu.$ After an arbitrary time $t,$ the locations of satellites after time $t$ are rotated by $\omega_st$ where $\omega_s$ is the angular speed of the LEO satellites. Based on the displacement theorem, the distribution of LEO satellites on each orbit  after the time $t$ still follows a Poisson point process of the same intensity $\mu$. Therefore, we conclude that the proposed satellite Cox point process $\Psi$ is time invariant. 
\end{IEEEproof}

\subsection{Number of Satellite Harvesters}
\begin{lemma}\label{lemma1}
	At any given time, the number of communicable orbits follows a Poisson random variable of mean ${\lambda\sin(\xi)}$ where $\xi $ is Eq. \eqref{azi_sensing}. 
	
	Further, the average number of the LEO satellites within the distance $\gamma $ is 
	\begin{equation}\label{L1}
		\frac{\lambda\mu}{\pi}\int_{0}^{\xi}\cos(\varphi)\arcsin(\sqrt{1-\cos^2(\xi)\sec^2(\varphi)})\diff \varphi.
	\end{equation} 
\end{lemma}

\begin{IEEEproof}
See Appendix \ref{A:0}
\end{IEEEproof}
\begin{figure}
	\centering
	\includegraphics[width=1\linewidth]{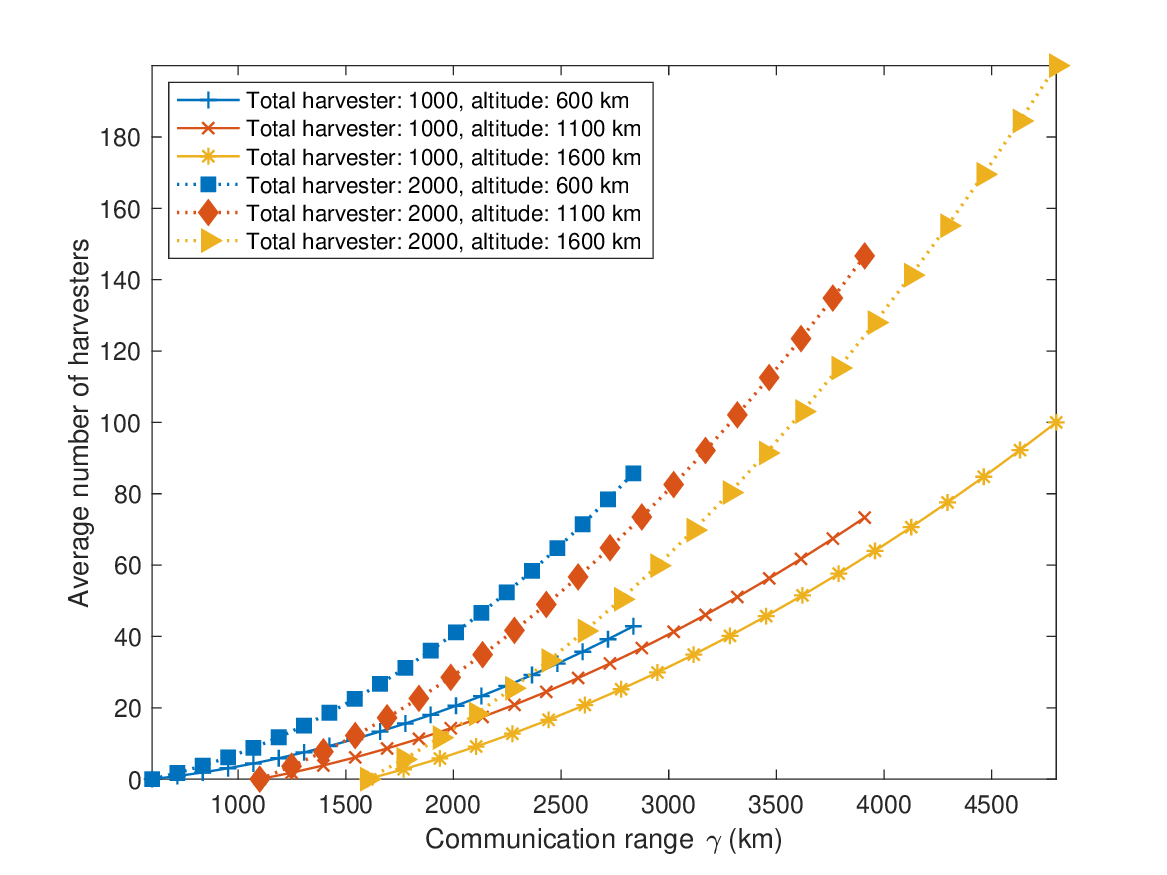}
	\caption{{The expected number of harvesters in the proposed network.}}
	\label{fig:expectednumberofharvesters}
\end{figure}
Fig. \ref{fig:expectednumberofharvesters} illustrates the expected number of LEO satellite data harvesters seen by the user at $(0,0,r_e)$ in the proposed network. We observe that the expected number of harvesters increases as the communication range increases,


As the value of $\gamma $ increases, the expected number of harvesters $\bE[N]$ also increases. Similarly, as the product of  parameters  $\lambda$ and $\mu$ (the total number of harvesters) increases, $\bE[N]$ also increases. Eq. \eqref{L1} suggests that $\bE[N]$ exhibits a linear relationship w.r.t. $\lambda\mu$, representing the total number of harvesters in the system. 


In Fig. \ref{fig:expectednumberofharvesters}, the range of $\gamma$ on the $x$-axis varies for different satellite altitudes due to the geographic limitation: $\gamma \in[r_a, \sqrt{r_o^2-r_e^2}]$; for instance, $\sqrt{7500^2-6400^2}=3900$ km.

	Suppose $\gamma = 1000$ km and $r_a=600 $ km, or equivalently $r_o = 7000$ km.  Using Eq. \eqref{L1}, we get $\bE[N] \approx  0.0036\lambda \mu $, where $\lambda\mu$ is the average number of LEO satellites. If  $\lambda\mu=1000,$ there are approximately $3.6$ satellites within distance $\gamma$ from the typical user. Provided that there are $1000$ LEO satellites,  $3.6$ is not a very high number. On the other hand, using $\gamma =2000$ km, we have $\bE[N]\approx 0.02\lambda\mu$. Therefore, with $\lambda\mu = 1000,$ we get $\bE[N] \approx 20.$ Note that as a single number, the variable $\gamma$ captures various network constraints including the transmit and receive antenna array patterns, the transmit power, or thermal noise of LEO satellites. 

\section{Harvest Time Fraction}
The harvesting capability of the proposed system is influenced by both the number of LEO satellites and the fraction of time that the typical user has access to a LEO satellite for communication. Leveraging the time-invariant property described in Lemma \ref{P1}, we can now analyze the fraction of time that the typical user can find any satellite within $\gamma$. This fraction of time corresponds to the probability that the typical user can find at least one LEO satellite within a distance of $\gamma$ at any given time. As a result, the metric time fraction takes a value between $0$ and $1$.  

\begin{theorem}\label{T:1}
	The harvest time fraction, i.e., the fraction of time that there is at least one LEO satellite within range $\gamma $ is 
	\begin{align}
		\cF &= 1- e^{-{\lambda}\int_0^{\xi} \cos(\varphi)\left(1- e^{-\mu/\pi \arcsin(\sqrt{1-\cos^2(\xi)\sec^2(\varphi)})}\right)\diff \varphi }.
	\end{align}
\end{theorem}
\begin{IEEEproof}
See Appendix \ref{A:1}.
\end{IEEEproof}


\begin{figure}
	\centering
	\includegraphics[width=1\linewidth]{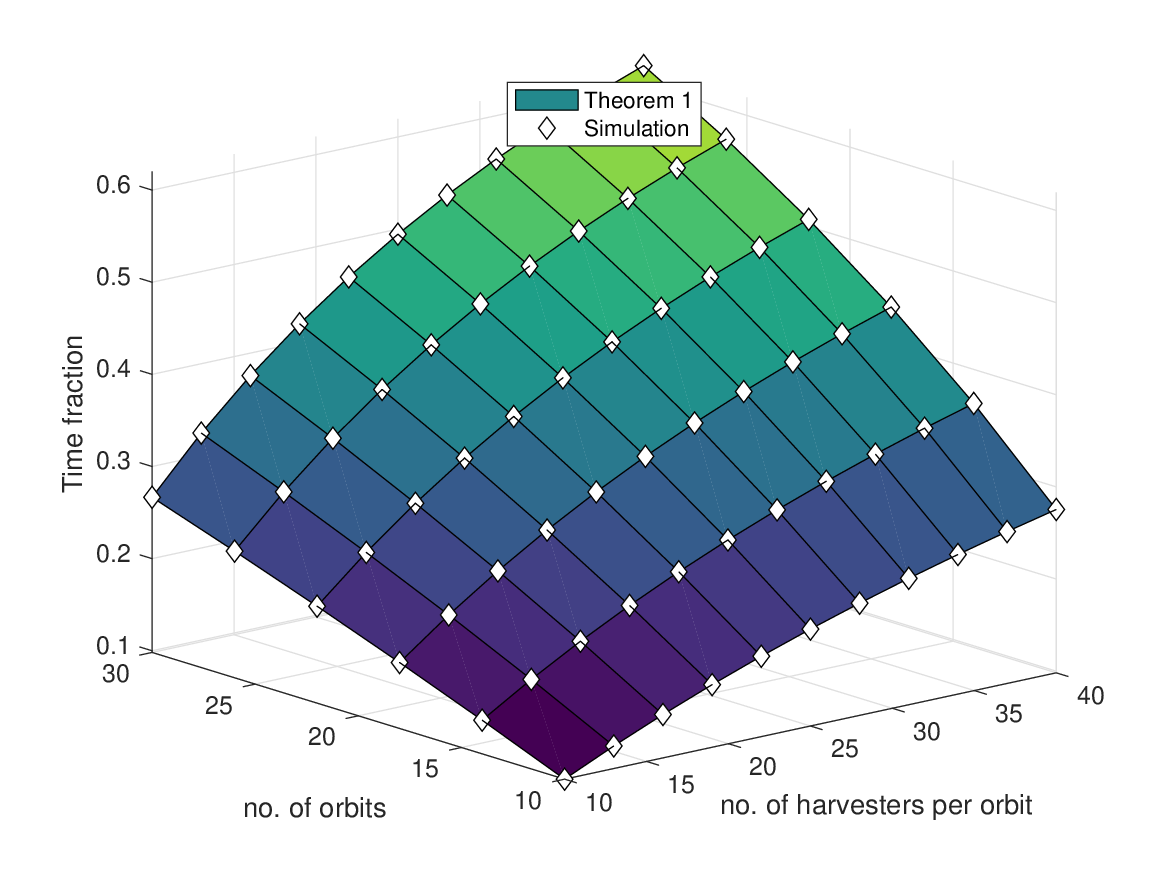}
	\caption{The time fraction of the user. The harvester altitude is $700$ km and the maximum communication distance is $750 $ km. Simulation results validate the accuracy of the analytic formula presented in Theorem \ref{T:1}.}
	\label{fig:timefractiongamma700}
\end{figure}

\begin{figure}
	\centering
	\includegraphics[width=1\linewidth]{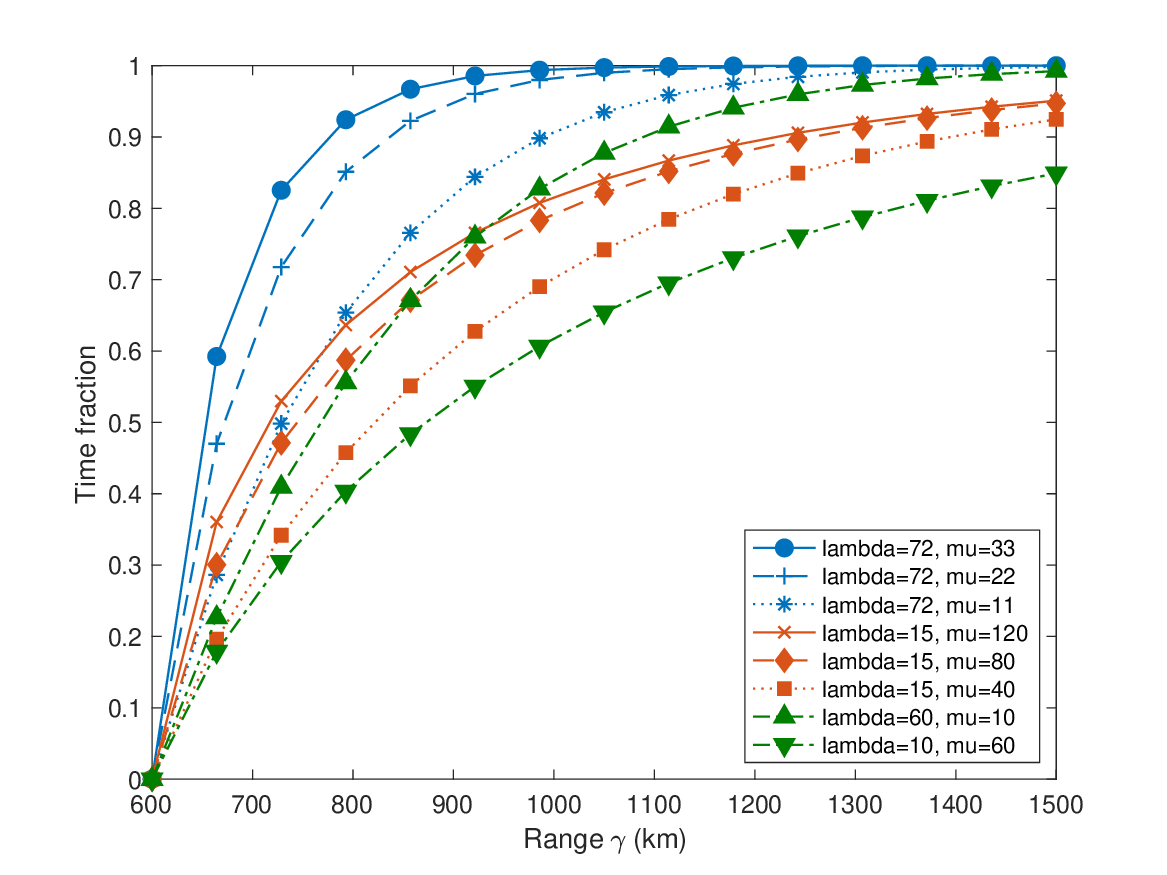}
	\caption{The time fraction  in Theorem \ref{T:1} for various numbers of satellites.}
	\label{fig:gammavariablet1}
\end{figure}
	The derived time fraction is the time average that the typical user finds any LEO satellite within distance $\gamma$. In other words, over a long period of time $T$, the amount of time that the typical user finds any LEO satellite harvester is given by the product $T \times \cF$ representing the total time and time fraction, respectively. Roughly speaking, if $\cF= 0.8, $ the typical user is expected to have at least one LEO satellite harvester within distance $\gamma$ at $80$ time units (slots) out of every $100$ time units (slots). 
	
Moreover, due to the rotation invariant property of the satellite point processes, the harvest time fraction of the typical user statistically \emph{represents} the harvest time fractions of all users in the network.

A notable observation is that Theorem \ref{T:1} is independent of the satellite speed factor $\omega_s$. This is because the satellites on each orbit form a Poisson point process, and once their locations are initialized, they move at the same speed on the orbit. Therefore, the speed factor $\omega_s$ does not affect the analysis. On the other hand, the time fraction is influenced by distributional parameters such as the number of orbits $\lambda$, the number of satellites per orbit $\mu$, the communication range $\gamma$, and the satellite altitude $r_a$. Fig. \ref{fig:timefractiongamma700} depicts the harvest time fraction as a function of $\lambda$ and $\mu$. We confirm that the simulation results confirm the analysis.

Fig. \ref{fig:gammavariablet1}  illustrates the time fraction w.r.t.  $\gamma$. We observe that a higher $\gamma$ leads to a greater time fraction for the typical user, simply because there are more satellites within a distance of $\gamma$ from the user. {It is worth noting that under the proposed network model, the time fraction is influenced by both the number of orbits and the number of satellite harvesters per orbit. For example, the time fractions for the cases with $\lambda=15, \mu=120$ and $\lambda=72,\mu=22$ are significantly different, despite both cases having a similar total number of satellite harvesters. Theorem \ref{T:1} accounts for the behavior of the time fraction as an explicit function of $\lambda$, $\mu$, and $\gamma$.}  

In practical data harvesting architectures, LEO satellite network operators may need to optimize their resources and allocate their finite number of LEO satellite harvesters wisely to build an efficient data harvesting architecture. The aforementioned result in Theorem 1 indicates that increasing the number of orbital planes is more beneficial than increasing the number of LEO satellites per plane in terms of maximizing the harvest time fraction. Specifically, with the same number of $600$ satellite harvesters, a dense orbit scenario with $\lambda=60,\mu=10$  shows approximately $30\% $ to $50\%$ higher harvest time fraction compared to a sparse orbit scenario with $\lambda=10$, $\mu=60$. Similarly, to achieve the same harvest time fractions, network users with $\lambda=60$ can have about $10$\% to $20$\% smaller $\gamma$ compared to users with $\lambda=10$. As discussed, the value of $\gamma$ aggregates various non-geometric network aspects of data harvesting architectures. If $\gamma$ is determined by the energy detection threshold of LEO satellite harvesters in practice, this implies that all users in a dense orbit scenario can effectively reduce their transmit powers by a factor of $(1+\frac{20\%}{100})^{\alpha}$ while achieving the same energy detection threshold. Consequently, in terms of overall energy efficiency, this suggests that data harvesting architectures with constellations consisting of many orbits substantially reduce the total energy consumption for data harvesting.


While the derived harvest time fraction provides information on the availability of LEO satellite harvesters, it does not directly indicate the amount of data that the satellite harvesters can actually collect from network users. In the following section, by utilizing the unified framework developed in this paper, we investigate the harvesting capability of the proposed system by taking into account the random dynamics of communication links from users to satellite harvesters.
	\begin{figure}
	\centering
	\includegraphics[width=1\linewidth]{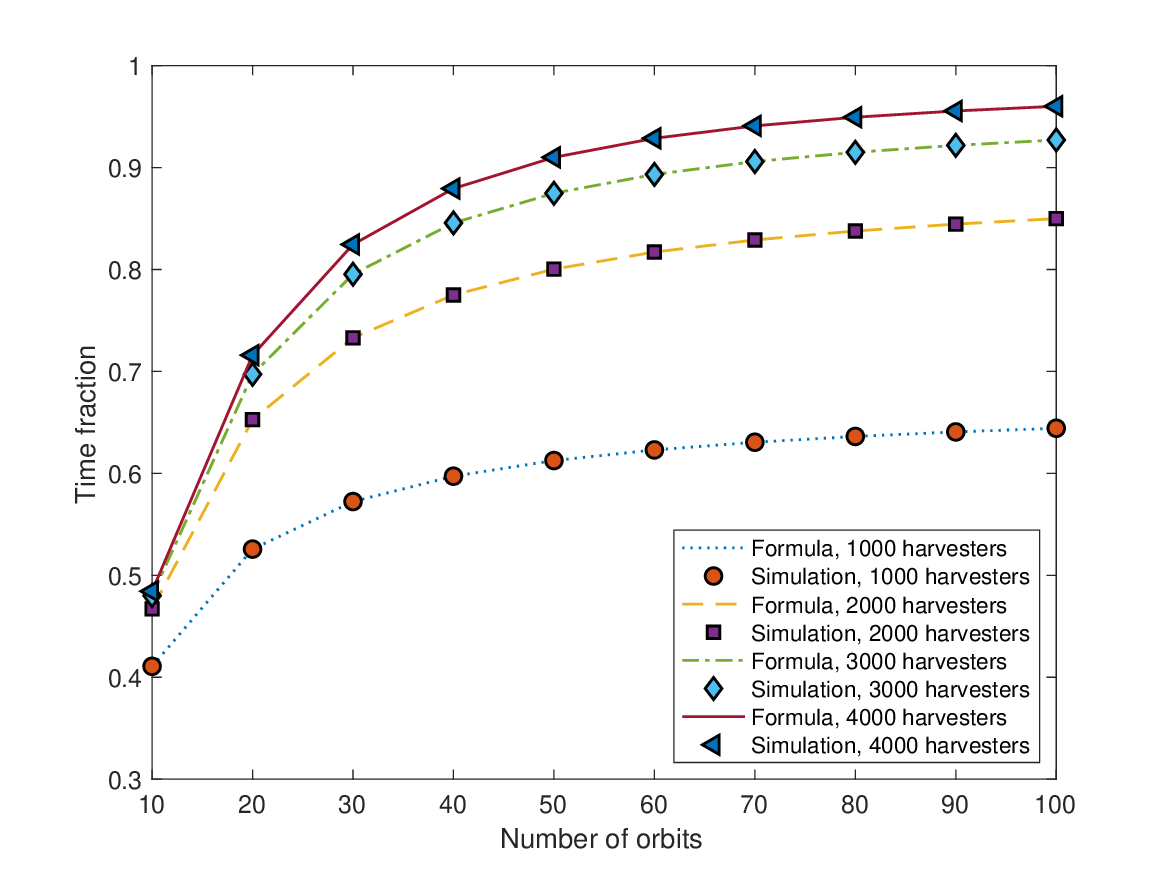}
	\caption{{Time fraction for given numbers of harvesters. We set $\lambda\mu=\{1000,2000,3000,4000\}$.}}
	\label{fig:theorem1optimization}
\end{figure}	
\begin{example}[Optimizing number of orbital plane] Fig. \ref{fig:theorem1optimization} illustrates the time fraction of the network while keeping the number of satellite harvesters constant. Given that the number of satellite harvesters is determined by the product of the number of orbital planes and the number of satellites per plane, this figure provides valuable insights into optimizing the number of orbital planes for a given number of satellite harvesters. For all values of $\lambda\mu={1000,2000,3000,4000}$ we observe a consistent trend: as the number of orbital planes increases, the time fraction consistently rises. This observation suggests that in deployment scenarios where the total count of satellites is limited, increasing the number of orbital planes is the optimal strategy for improving data harvesting time fraction, as opposed to increasing the number of satellites within each orbital plane.

\end{example}

\section{Data Harvest per Pass}

This section focuses on the analysis of the harvesting capability of the proposed delay-tolerant network, particularly when the numbers of satellite harvesters and of the users is low. Specifically, we evaluate the amount of data that a typical user can upload during a satellite passage, measured in bits. This metric represents the expected amount of data that each ground user can upload to a moving satellite while the satellite is within a distance of $\gamma$. {We assume that a sufficient number of ground-based gateways are available so that the data collected by satellite harvesters can be delivered to the gateways without network congestion \footnote{To fully assess the impact of gateway distribution on harvesting capacity, one needs to understand the joint distributions of satellite harvesters, user distributions, and gateways. This is outside the scope of this paper}.}

To highlight the motion of satellites in the analysis, we assume that when a satellite harvester appears at a distance of $\gamma$ from the typical user, the user starts uploading its data to the satellite harvester and continues uploading until the harvester goes out of range. Consequently, the obtained value reflects the amount of data that each user can upload while being served by a single satellite harvester.

This association is also considered to account for a practical limitation. When the number of satellite harvesters is low, it may be infeasible to establish new connections with other satellites (as indicated in Lemma \ref{P1} and the provided example). Additionally, even if new connections are established, they do not always guarantee maximum data harvest. In scenarios with a higher number of satellite harvesters, we will introduce and utilize the harvesting capacity in Section \ref{S:harvestcapacity} to further analyze the system's performance.

\begin{figure*}
	\begin{align}
		&\frac{mB_w}{\omega_s}\int_{0}^{\omega_0}  \overbar{F_H}\left(\frac{\tau \sigma^2}{pg} (r_o^2-2r_or_e\cos(\omega)\sin(\phi)+r_e^2)^{\alpha/2}\right)\diff \omega,\label{eq:Theorem1}\\
		&\frac{B_w}{\omega_s}\int_{0}^\infty \int_{0}^{\omega_0}  \overbar{F_H}\left(\frac{\sigma^2(2^u-1)}{pg} {(r_o^2-2r_or_e\cos(\omega)\sin(\phi)+r_e^2)}^{\frac{\alpha}{2}}\right)\diff \omega \diff u.\label{eq:Theorem1-2}
	\end{align}
	\rule{\linewidth}{0.1mm}
\end{figure*}
\begin{theorem}\label{T:2}
	Suppose a fixed ${m}$-ary modulation and associated SNR threshold $\tau$. For orbit $l(\theta,\phi)$, the average successful data harvest per pass from a satellite on this orbit is given by Eq. \eqref{eq:Theorem1}
	where $\overbar{F_H}(x)$ is the CCDF of the random variable $H$ in Eq. \eqref{CDF of H}. 
	
	Suppose an adaptive modulation where modulation rate is determined by the SNR, i.e., $m=\log2(1+\SNR)$. Then, for orbit $l(\theta,\phi)$, the harvest data per pass from a satellite on this orbit is given by Eq. \eqref{eq:Theorem1-2}. 
\end{theorem}
\begin{IEEEproof}
	See Appendix \ref{A:2}.
\end{IEEEproof} 

\begin{figure}
	\centering
	\includegraphics[width=1\linewidth]{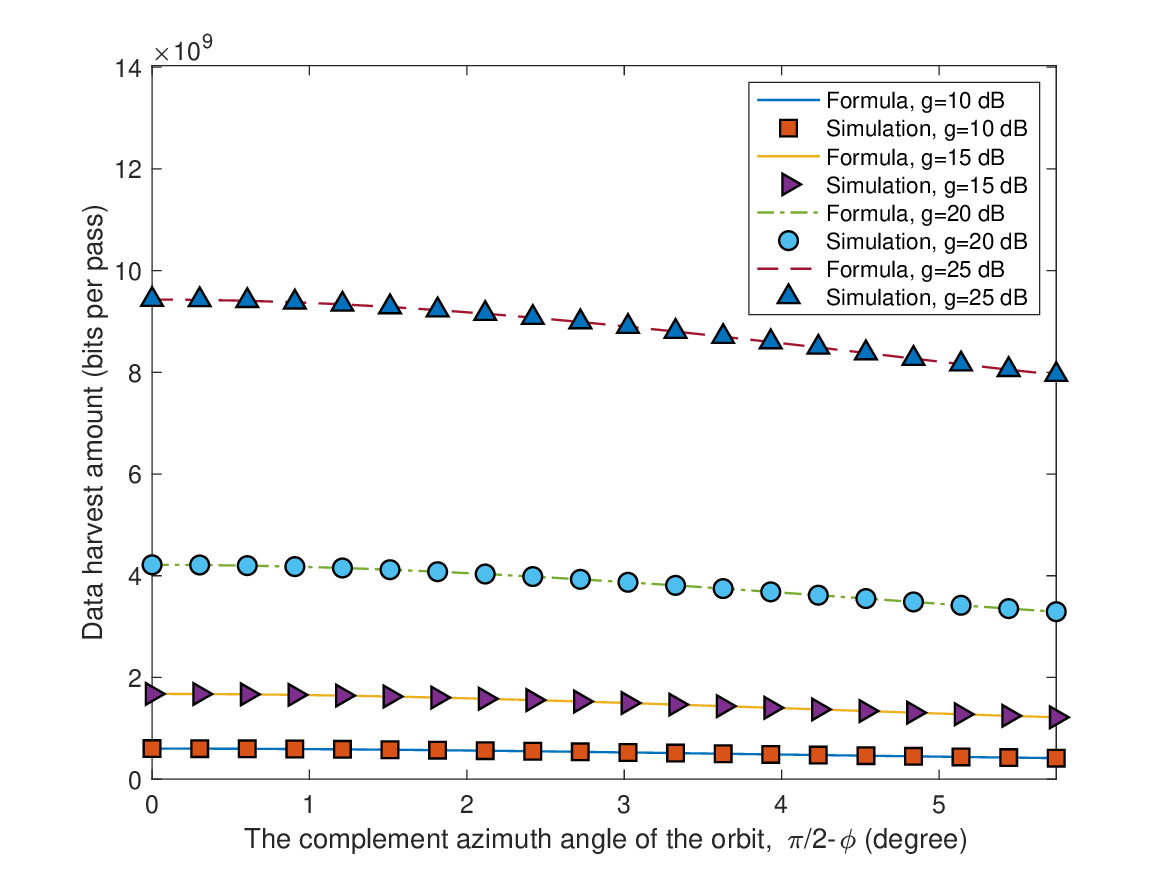}
	\caption{{Data harvest in Theorem \ref{T:2} with $m=1$. We use values in Table \ref{Table:2}. Simulation results validate the accuracy of the analytic formula presented in Theorem \ref{T:2}.}} 
\label{fig:theorem2}
\end{figure}

In Theorem \ref{T:2}, we evaluate the average amount of harvest data under the assumption that once each user finds its satellite within range, it continues uploading its data until the satellite goes out of range. This assumption is practical when the number of satellites is small, resulting in a low probability of having more than one satellite within the user's range.
\begin{table}
\centering
\caption{{Network simulation parameter}}\label{Table:2}
\begin{tabular}{|c|c|}
	\hline
	Variable & Value \\
	\hline
	Transmit power $p$& $30$ dBm \\
	\hline
	Thermal noise  & $-174$ dBm/Hz \\
	\hline
	Bandwidth $B_w$ & $20$ MHz \\
	\hline
	Carrier frequency & $1$ GHz \\
	\hline
	Aggregate antenna gain $g$& $\{10,15,20,25\}$ dB \\
	\hline
	Satellite harvester altitude & $600$ km  \\
	\hline 
	Max comm. distance $\gamma$& $900$ km  \\
	\hline 
	Path loss exponent $\alpha$& $2$\\
	\hline 
\end{tabular}
\end{table}
Fig. \ref{fig:theorem2} shows the data harvest per pass in the proposed network model. Table \ref{Table:2} summarizes network parameters to produce the results. We confirm that the simulation results confirm the derived formula. We show that the amount of data harvest is the maximum when the satellite passes directly above the typical user, i.e., $\phi=\pi/2$. In this case, the time duration of the satellite passing the typical user is the longest and the average link distance is the shortest among all other satellites on different orbits. On the other hand, the amount of harvested data is the lowest when the orbit's inclination is close to $\pi/2-\xi$ or $\pi/2+\xi$. 

In practical delay-tolerant networks, the elevation angle of a passing satellite for users at a certain location may be known to the satellite network operator. If such an angle is known, the data harvest per pass can be precisely evaluated by referring to the azimuth angle of the $x$-axis in Fig. \ref{fig:theorem2}. In cases where the elevation angles of satellite harvesters are unknown,  Theorem \ref{T:2} and Fig. \ref{fig:theorem2} can be still utilized to estimate the total data harvested from such satellites. These projections or estimations of data harvest per pass from a given satellite constellation can serve as useful guidelines to analyze the performance of data harvesting from satellites in space. The $x$-axis of Fig. \ref{fig:theorem2} is the complement azimuth angle, namely $\pi/2-\phi$ where $\phi$ is the orbit inclination. 
\begin{remark}
In the derivation of the data harvester per pass, we assume that the typical user maintains its uplink connection to the association harvesters until it is out of communication range. Therefore, the satellite harvester collects data only from the target ground user, and there is no uplink interference from nearby ground users. This data harvesting approach is suitable for areas with low numbers of satellite harvesters and ground users. In the rare case where a satellite harvester identifies more than one ground user for a given area, only one of those ground users is allowed to communicate with the satellite harvester. The rest of the ground users are configured to be served by the next cycle of the satellite harvester or other harvesters covering the same area. In the next section, we will address data harvesting in scenarios where the numbers of satellite harvesters or ground users are high.
\end{remark}



\section{Harvesting Capacity}\label{S:harvestcapacity}


This section considers a scenario where the number of satellite harvester is high and in this case, the typical user is assumed to communicate with its closest satellite amongst all satellites within distance $\gamma$. Note that as satellite harvesters move following their orbits, the spatial distribution of satellite harvesters changes and so do the distances between the harvesters and the typical user. Therefore, to incorporate this redistribution of harvesters, we assume that at any given time, the typical user establishes its uplink connection to its nearest satellite harvester. This approach ensures that the typical user is associates with the satellite providing the maximum average power, which leads to a larger harvesting capacity. 

This section defines the harvesting capacity of the proposed network architecture as the average achievable rate of the typical uplink communication from the typical user to its nearest satellite harvester within distance $ \gamma $. 

\begin{figure*}
		\begin{align}
		\frac{\lambda\mu B_w}{\pi r_er_o}\int_{0}^\infty \int_{r_a}^{\gamma}\overbar{F_H}\left(\frac{(2^r-1)u^\alpha}{pg/\sigma^2}\right) \exp{\left(-{\lambda}\int_0^\kappa \cos(\varphi)\left(1-\bar{g}(\varphi,u)\right)\diff \varphi\right)}\left(\int_0^\kappa \frac{  \bar{g}(x,u)}{\sqrt{1-\cos^2(\kappa)\sec^2(\varphi)}}\diff x\right)\diff u \diff r.\label{eq:Theorem3}
	\end{align}
		\rule{\linewidth}{0.1mm}
\end{figure*}
\begin{theorem}\label{T:3}
	The harvesting capacity of the proposed network model is given by Eq. \eqref{eq:Theorem3} 
where $pg/\sigma^2$ is the received SNR at 1 meter, $\overbar{F_H}(x)$ is the CCDF of $H,$ and $\bar{g}(x,u)$ is Eq. \eqref{30}, respectively. 
\end{theorem}
\begin{IEEEproof}
	See Appendix \ref{A:3}.
\end{IEEEproof}
%
%
\begin{figure}
	\centering
		\includegraphics[width=1\linewidth]{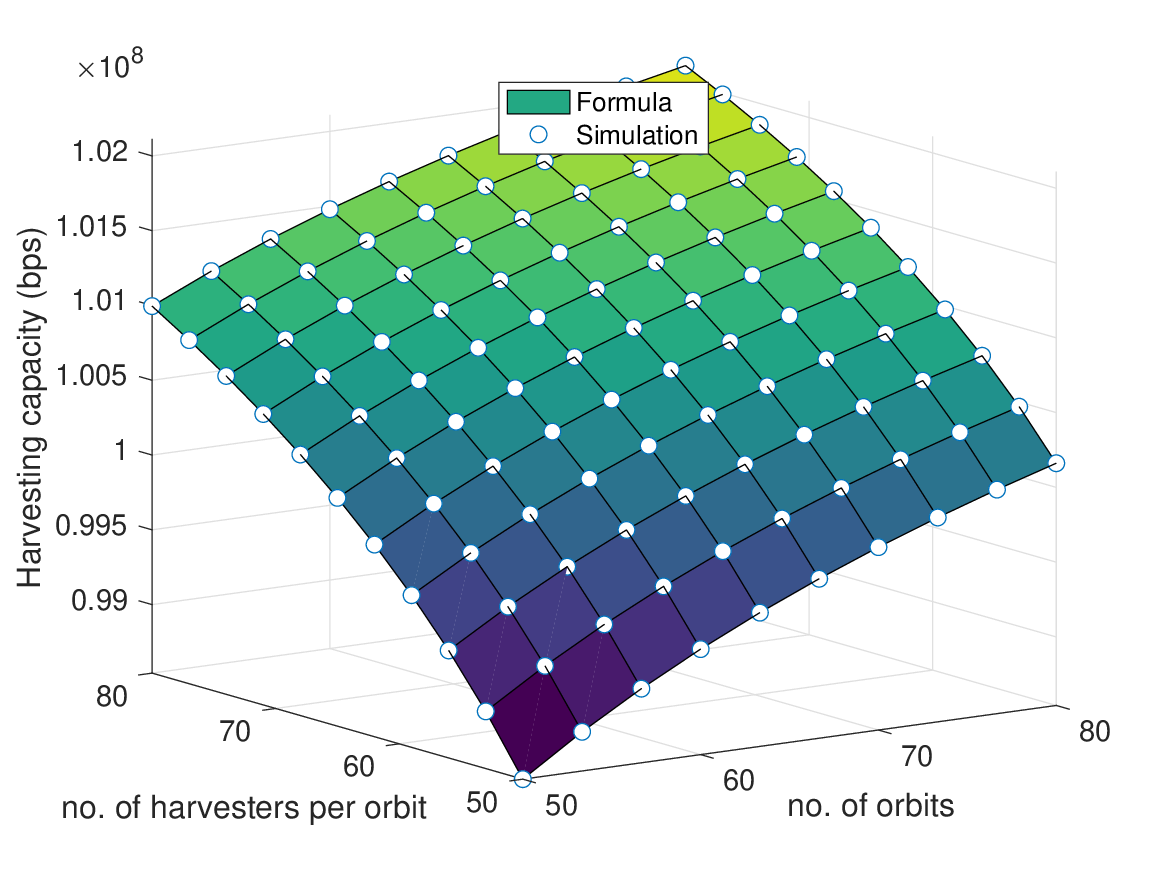}
	\caption{{Harvesting capacity. Simulation results validate the accuracy of the analytic formula presented in Theorem \ref{T:3}.}} 
	\label{fig:theorem3harcapgamma1200}
\end{figure}

\begin{remark}
In this section, we evaluate the harvesting capacity, particularly in scenarios involving numerous satellite harvesters and users. For each satellite harvester, there may be multiple ground users associated with it, and we assume that such network users are allowed to engage in time-sharing, ensuring that only one network user can upload its data to its associated satellite harvester at any given time \cite{10209551}. In the rare case where the coverage of many satellite harvesters are concentrated in a small area, we assume that such neighboring satellites engage in coordinated frequency resource sharing among themselves through inter-satellite links, thereby minimizing inter-satellite interference. \cite{7466793,9852737,10209551}
\end{remark}

Fig. \ref{fig:theorem3harcapgamma1200} shows the harvesting capacity obtained by our analytical formula and by Monte Carlo simulation experiments. For the analysis, we use $\gamma=1200$ km and $g=35$ dB, and values in Table \ref{Table:2}.  As the number of orbits or the number of harvesters per orbit increases, the harvesting capacity monotonically increases. In the figure, both parameters grow from $50$ to $80$. By leveraging the formula in Theorem \ref{T:3}, one can avoid spending a significant amount of time and computation costs to get the harvesting capacity of the delay-tolerant satellite networks. 

With the current values of network parameters, the harvesting capacity increases as $\lambda$ the number of orbits or $\mu$ the number of harvesters per orbit increases. Please note that adding more orbital planes for harvesters leads to a larger harvesting capacity. This is primarily because as there are more orbital planes, the typical user is likely to upload more data to the association satellite at a closer distance.  

In practice, the harvesting capacity provides valuable insights for network operators regarding the average harvest capacity when enabling satellite constellations to collect data from surface devices on Earth. By utilizing Theorem \ref{T:3}, network operators can assess the feasibility and quality of the services they can offer to their ground subscribers, as well as determine the necessary frequency and time resources to enable such service. Additionally, the harvesting capacity allows for the estimation of the maximum data rate that each user can generate and transmit to the satellite harvesters while avoiding excessive data congestion at the terminals. A detailed discussion is left for future work.

\begin{table}\centering
	\caption{Theorems \ref{T:2} and \ref{T:3}} \label{Table:1}
	\begin{tabular}{|c|c|c|}
		\hline
		& Theorem \ref{T:2} & Theorem \ref{T:3}\\ 
		\hline
		Metric	&  Amount of harvest per pass & Harvesting capacity \\
		\hline
		Unit	&  Bits per pass & Bits per sec\\
		\hline
		Assoc. harv.  &  Harvester on $l(\theta,\phi)$ & Nearest harvester\\
		\hline
		Harvester density  &  Low   &  Mid to high\\
		\hline
	\end{tabular}
\end{table}

\begin{remark}[Difference of data harvest and harvesting capacity]
	It is worth noticing the difference between Theorem \ref{T:2} and Theorem \ref{T:3}. First, Theorem \ref{T:2} shows the amount of data successfully picked up by a single satellite harvester from a single network user, as the satellite progresses within the communication range of the single network user. The associated uplink from the typical user to its satellite is maintained  as long as the satellite is within the range. Although this metric is useful for predicting the amount of uploaded data for remote or disconnected devices as a function of each satellite pass, it does not directly represent the optimal data harvesting capacity of the delay-tolerant satellite network. On the other hand, Theorem \ref{T:3} shows the amount of data picked up by the satellite harvester constellation when a typical user is assumed to upload its data to its nearest LEO satellite. Since the metric exploits the shortest link between users and LEO satellites and considers their relative motion too, the derived harvesting capacity is useful to predict the optimal harvesting capability of a given LEO satellite constellation with LEO satellite harvesters.  Theorem \ref{T:2} is useful especially when the number of satellite harvesters is small and thus the data harvesting is opportunistic; in contrast, Theorem \ref{T:3} is applicable to any constellation with a large number of satellite harvesters. In this paper, we give both metrics to provide a comprehensive analysis of the harvesting capability of the proposed network model. Table \ref{Table:1} summarizes the key differences between Theorems \ref{T:2} and \ref{T:3}.
\end{remark}

\section{Delay Distribution}
In a data harvesting architecture leveraging moving LEO satellites, the system's delay consists of two main components: the delay from the terminal user to the LEO satellite and the delay from the associated LEO satellites to the ground infrastructure. Especially when the number of available LEO satellites is small, or when the communication range $\gamma$ is small, the first delay component dominates the second delay component since it may take a significant amount of time for a terminal user to locate a LEO satellite within $\gamma$. Tthe total end-to-end delay in the harvesting architecture is mainly influenced by the time that users need to wait until they find LEO satellites to upload their data.

\begin{figure}
	\centering
	\includegraphics[width=0.7\linewidth]{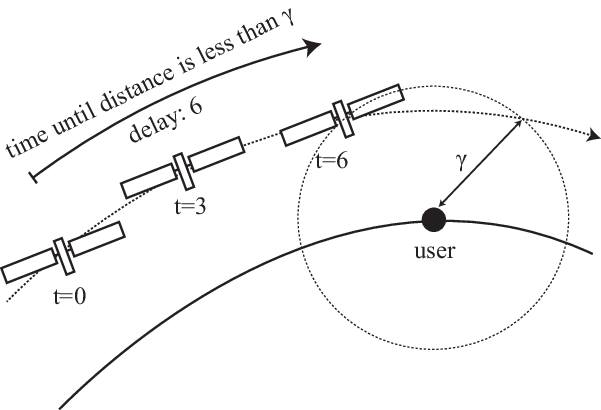}
	\caption{After a delay of $t=6$ time units, the LEO satellite is within distance $\gamma$ from the user.}
	\label{fig:conceptharvesting}
\end{figure}

This section focuses on the aforementioned delay, specifically the amount of time that a typical user must wait until it finds a LEO satellite at a distance less than $\gamma$. We exclude non-geometric or controllable aspects of delays, such as the scheduling delay for certain users to be allocated for uplink communications or the inter-satellite delay from LEO satellites to other LEO satellites to the ground gateway infrastructure. These controllable delays can be managed and improved by network operators \cite{8473415,9755278,9970355}. Fig. \ref{fig:conceptharvesting} illustrates the delay examined in this paper. It shows the typical user upload its data to the satellite at time $t=6$, resulting in a delay of $6$ time units.

\par Before analyzing delay distribution, let us first define two extreme values delay takes, zero or infinite. The delay is zero if there is at least one LEO satellite within range $\gamma$ all times. On the other hand, the delay is infinite if there is no satellite within range $\gamma$ for all time. For instance, this event occurs if $|\phi-\pi/2|>\xi.$

\begin{corollary}\label{cor:1}
	$\bP(D=0) $ the zero delay probability is given by 
	\begin{equation} 
	 1- e^{-{\lambda}\int_0^{\xi} \cos(\varphi)\left(1- e^{-\frac{\mu}{\pi} \arcsin(\sqrt{1-\cos^2(\xi)\sec^2(\varphi)})}\right)\diff \varphi }.\nnb
	\end{equation}
On the other hand, the delay is infinite with probability 
\begin{equation}
	\bP(D=\infty) = \exp\left(-{\lambda \sin(\xi) }\right).
\end{equation}
\end{corollary}

\begin{IEEEproof}
	The zero delay probability is obtained simply. The delay equals to zero if there is at least one LEO satellite within distance $\gamma $ from the typical user. Therefore, the nonzero delay probability is the probability that there is no LEO satellite at distance less than $\gamma. $ Revisiting the proof in Theorem \ref{T:1}, we get the result. 
	
	The delay is infinite if there is no orbit with inclination $\phi \in (\pi/2-\xi,\pi/2+\xi).$ Since the orbit process is a Poisson point process of density $\lambda\sin(\phi)/(2\pi)$ on $\cR,$ hence we arrive at $\bP(D=\infty) = \exp\left({-{\lambda\sin(\xi) }}\right).$
\end{IEEEproof}

\begin{figure*}
		\begin{equation}
		\bP( D < d )  = 1- \exp\left(-{\lambda}\int_{0}^{\xi} \cos(\varphi)\left(1 - e^{-\frac{\mu}{2\pi}\left(2\arcsin\left(\sqrt{1-\cos^2(\xi)\sec^2(\varphi)}\right)+\omega_s d \right)}\right)\diff \varphi\right).\label{eq:theorem4}
	\end{equation}
	\rule{\linewidth}{0.1mm}
\end{figure*}
\begin{theorem}\label{T:4}
	The  delay distribution is given by Eq. \eqref{eq:theorem4}.
\end{theorem}

\begin{IEEEproof}
	See Appendix \ref{A:4}.
  \end{IEEEproof}
  
  \begin{figure}
  	\centering
  	\includegraphics[width=1\linewidth]{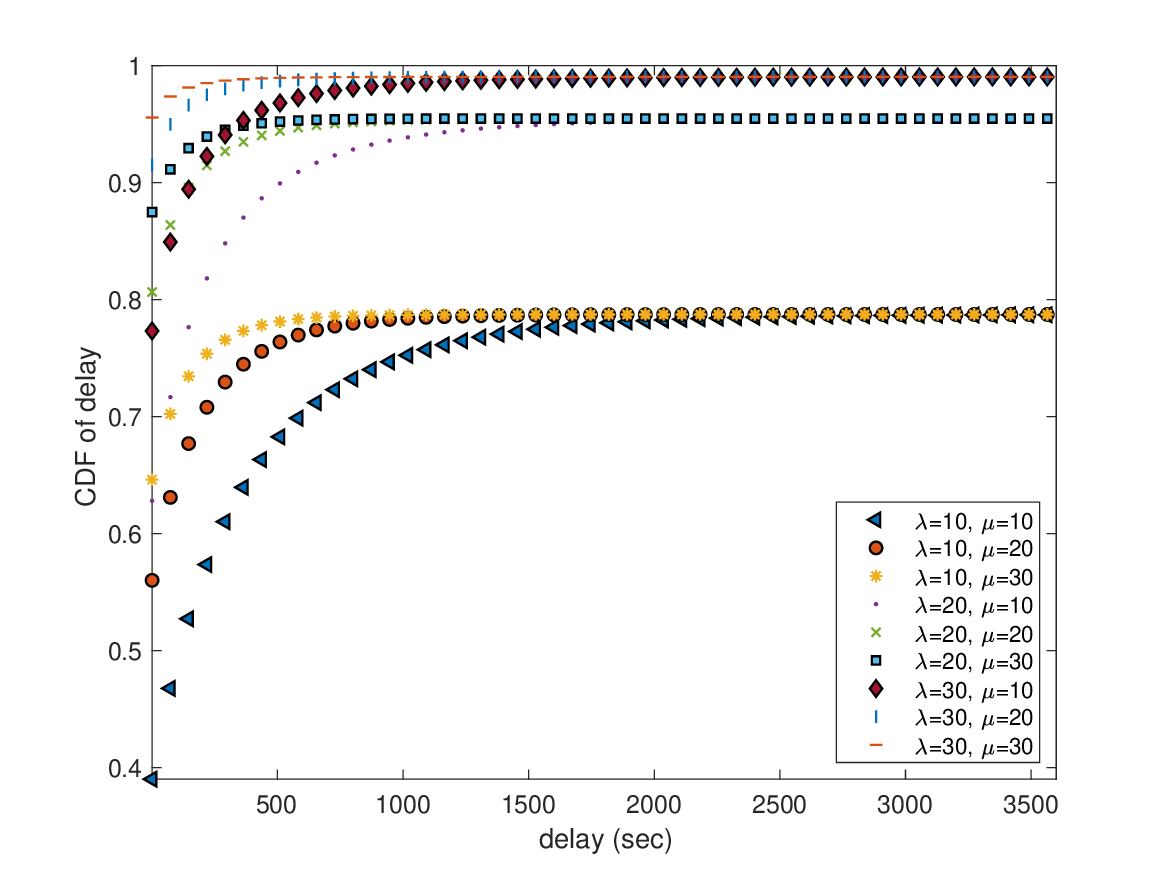}
  	\caption{Delay distribution of the proposed network model for $\lambda=10,20,$ and $30,$ and $\mu=10,20,$ and $30.$  }
  	\label{fig:theorem4delay}
  \end{figure}
  
Eq. \eqref{eq:theorem4} shows the delay distribution in the proposed network model as an explicit function of $\lambda,\mu,\omega_s,$ and $\gamma$.  Therefore, the impacts of network variables can be individually assessed from the formula. Note the above delay distribution is obtained by evaluating the delay distribution of the typical user located at $(0,0,r_e).$ Thanks to the joint rotation invariant property of the satellite point processes, Eq. \eqref{eq:theorem4} corresponds to the delay distribution averaged over all users in the network. 
  
Fig. \ref{fig:theorem4delay} shows the cumulative distribution function of the delay. It shows that there is certain mass at delay zero and specifically the delay is zero when there is at least one LEO satellite within distance $\gamma$ at time zero. For instance, the delay of the network with parameters $\lambda=10$ and $\mu=10$ is zero with probability $0.39.$  The delay of the network with $\lambda=30$ and $\mu=10$ is zero with probability $0.77. $ The figure also suggests that for a given value of $\lambda $, we have $\mu_1<\mu_2$ and then we arrive at $\bP(D_{\mu_1}<d)<\bP(D_{\mu_2}<d)$, i.e.,  in terms of the delay, a data harvesting architecture with the satellite density $\mu_1$ stochastically dominates the a data harvesting architecture with the satellite density $\mu_2$. From Fig. \ref{fig:theorem4delay}, we observe that the same number of satellite harvesters does not necessarily lead to the same delay distribution. For instance, although the constellations with $\lambda=20,\mu=10 $, and with $\lambda=10,\mu=20$ have the same number of satellite harvesters, their delay distributions are very different. 
 
 In practice, both Theorem \ref{T:4} and Fig. \ref{fig:theorem4delay}  can be utilized effectively to describe the benefits that network operators can expect by exploring various deployment scenarios, such as increasing the number of satellite harvesters or adding orbital planes.
 
 For instance, when obtaining the delay distribution of a LEO satellite network, a simulation-based analysis would require precise information on the exact number, locations, and movement directions of all satellites in the network. In contrast, the analysis presented in this paper only requires the number of orbital planes and the number of satellites, yielding the same statistical result. Moreover, the delay distribution provided in this paper is more valuable than simply evaluating the average delay of a system. It provides the tail distribution of the delay, which is a crucial factor in determining the performance of delay-tolerant networks.
 
 Additionally, in real delay-tolerant applications, the existence of a delay constraint is common, and this constraint can be adjusted depending on the choices made by network operators or the local density of users. In such cases, any topological update and its impact on the delay can be easily evaluated and assessed using Theorem \ref{T:4} and Figure \ref{fig:theorem4delay}. This enables network operators to test various deployment solutions and identify the optimal approach in a timely manner.

  

\section{Discussion}
Here, we compare the proposed Cox constellation to a polar constellation with a 90-degree inclination. To ensure a fair comparison, we develop the moment matching approximation method based on the work in \cite{choi2022analytical,choi2023Cox2}, matching the average numbers of orbits and satellite harvesters from different constellations.
  			
To determine the values for $\bar{\lambda}$ and $\bar{\mu}$ of the proposed Cox constellation, we leverage Lemma \ref{lemma1}. Let $\lambda$ and $\mu$ be the number of orbits and the number of satellites per orbit in the polar constellation. Then,  based on Lemma \ref{lemma1}, we get  
\begin{equation}
	\lambda = \bar{\lambda}\sin(\xi),\label{B}
\end{equation}
where the left hand-side is the mean number of visible orbits in the polar constellation and the right hand-side is the mean number of visible orbits in the proposed Cox constellation. 

Similarly, for both constellations, the mean numbers of satellites in the cap are given by  \begin{equation}
	\frac{\lambda\mu\xi}{\pi}=\frac{\bar{\lambda}\bar{\mu}}{\pi}\int_{0}^{\xi} \cos(v)\arcsin(\sqrt{1-\cos^2(\xi)\sec^2(v)})\diff v,\label{A}
\end{equation}
where the left hand-side is the mean number of visible satellite harvesters in the polar constellation and the right hand-side is the mean number of visible satellite harvesters in the proposed Cox constellation.  As a result, one use Eqs. \eqref{B} and \eqref{A} to determine $\bar{\lambda}$ and $\bar{\mu}$ the densities of the Cox point process based on the moment matching approximation. 

\begin{figure}
  		\centering
  		\includegraphics[width=1\linewidth]{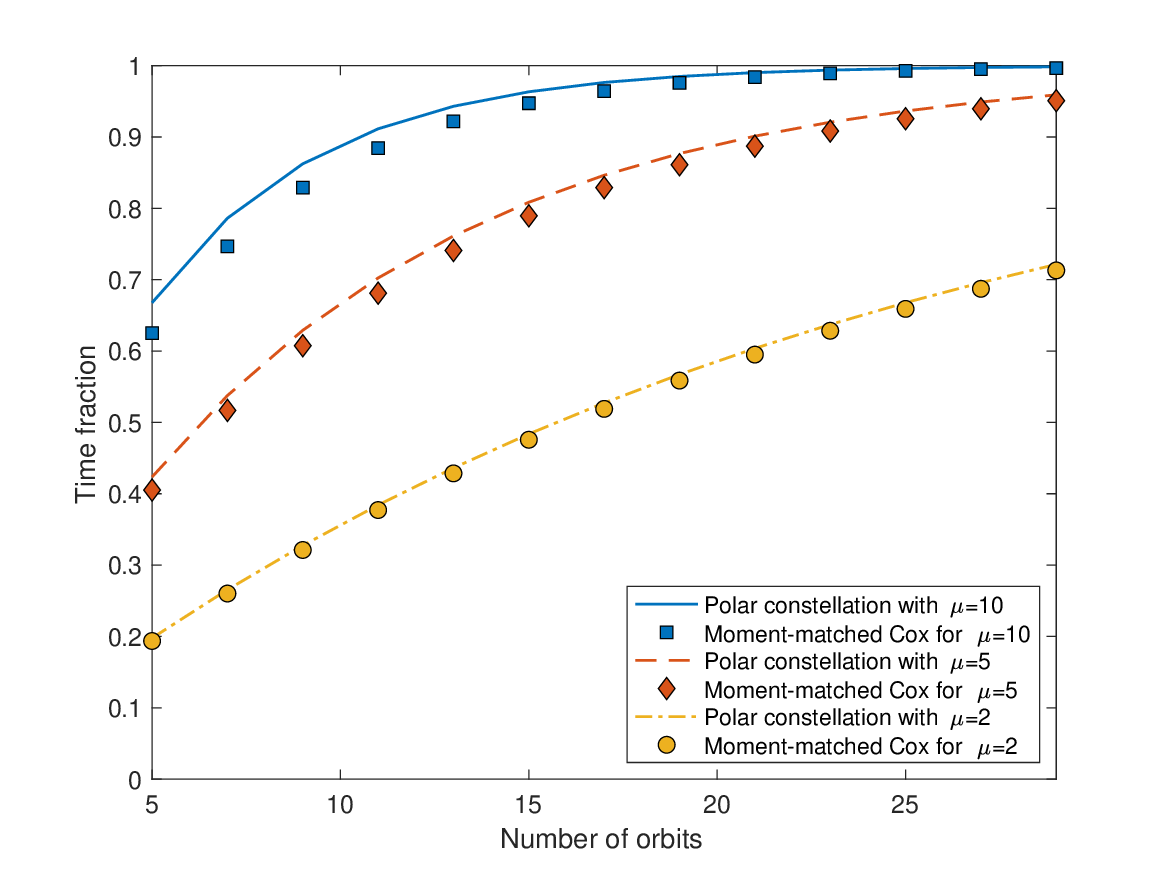}
  		\caption{{The time fractions of the proposed Cox constellation and the polar constellation. The satellite altitude is $1100$ km and the communication range is $1200 $ km.}}
  		\label{fig:theorem1compare}
  	\end{figure}

\begin{figure}
  		\centering
  		\includegraphics[width=1\linewidth]{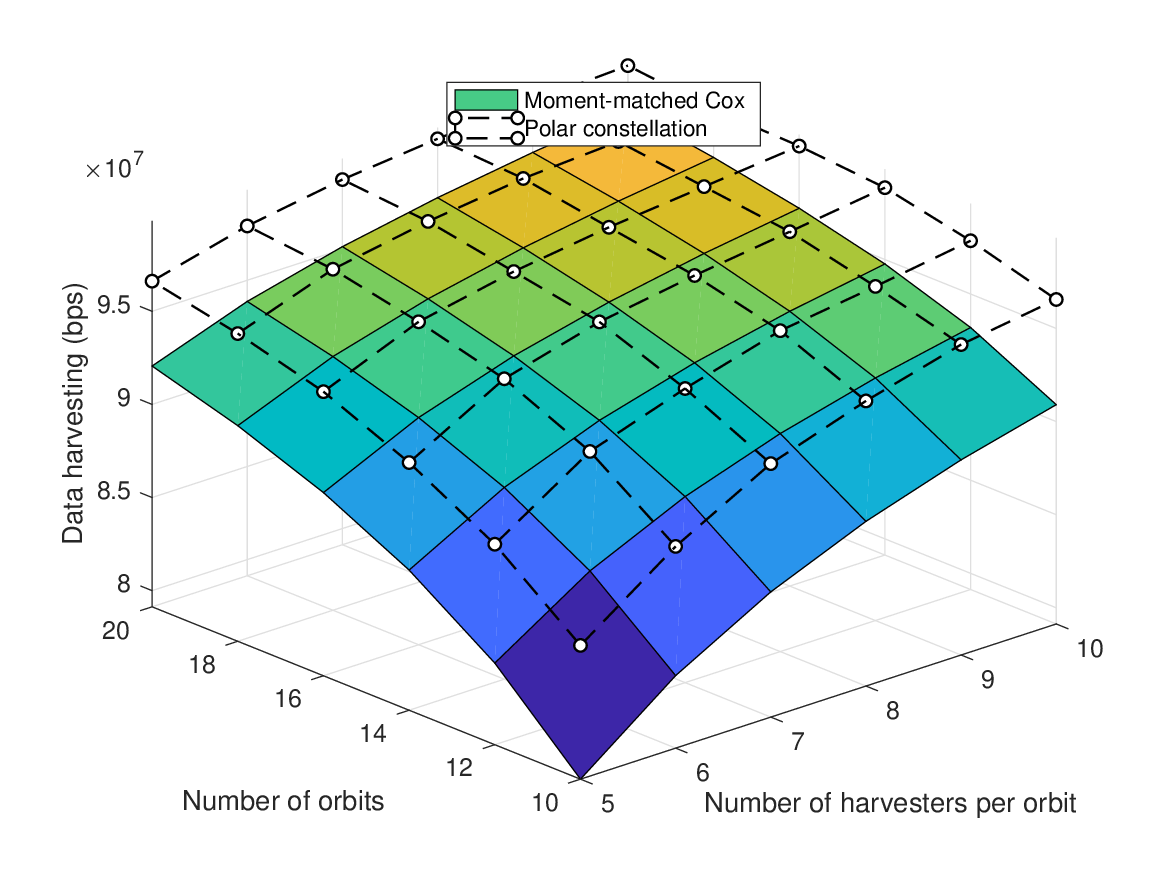}  		
  		\includegraphics[width=1\linewidth]{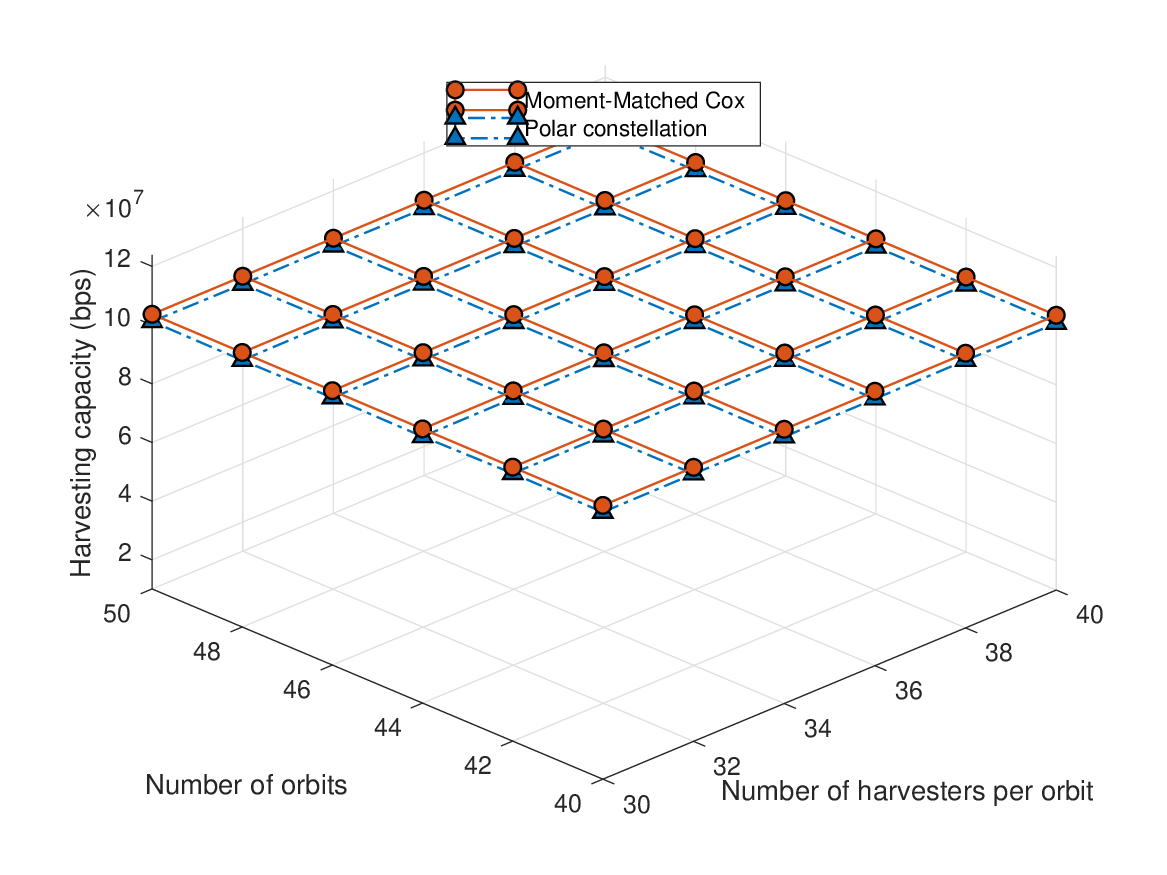} 
  		\caption{{The harvesting capacity of the polar and Cox constellations. We use the orbit altitudes $600$ km, $\gamma=1250$ km, and $g=35$ dB.}}
  		\label{fig:discussion2v1}
  	\end{figure}

\begin{figure}
  		\centering
  		\includegraphics[width=1\linewidth]{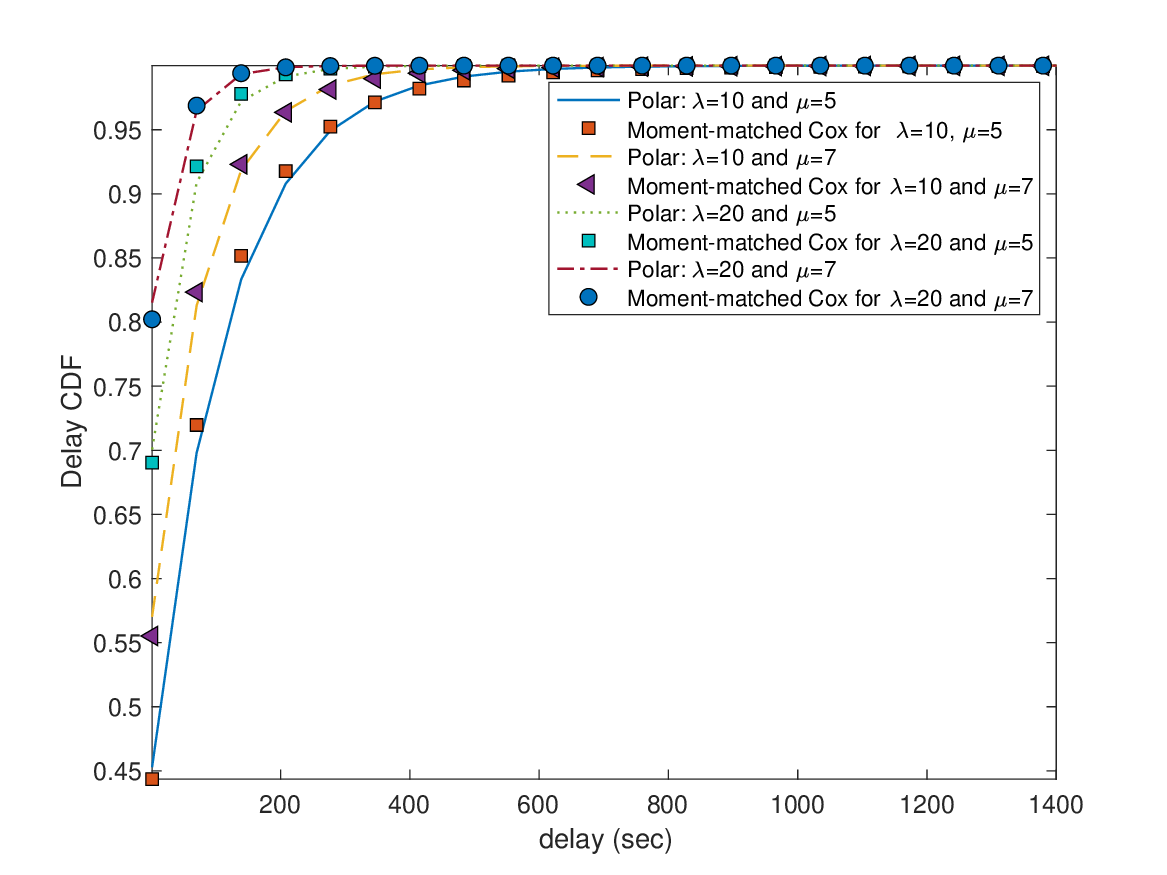}
  		\caption{{The delay distributions of the polar constellation and the proposed Cox constellation. The satellite altitude is 600 km and the communication distance is 650 km.}}
  		\label{fig:discussion3}
  	\end{figure}

In below, we determine $\bar{\lambda}$ and $\bar{\mu}$ based on the above moment matching approximation. Fig. \ref{fig:theorem1compare} displays the time fractions of the proposed network and a polar constellation with orbits having 90-degree inclinations. Fig. \ref{fig:discussion2v1} shows the harvesting capacity of the Cox and polar constellations, while Fig. \ref{fig:discussion3} presents the delay distributions. Worth noting is that, with the proposed moment matching method, we observe that the Cox constellation approximates the polar constellation with some errors. In addition, the delay distribution of the Cox constellation closely resembles that of the polar constellation, indicating the effective practical replication of the polar constellation by the proposed Cox configuration. Regarding the time fractions and delay distributions, the differences between the polar and Cox constellations are marginal, approximately less than $4\%$.
  			
  			In these figures, differences exist due to the fact that in the polar constellation, all orbits have a 90-degree inclination, whereas in the Cox constellation, inclinations are randomly distributed between 0 and 90. This is noticeable in Fig. \ref{fig:discussion2v1} where harvesting capacities are provided. It is worth noting that the difference is negligible for dense scenarios e.g., $\lambda > 40,\mu > 40$, which is the primary case that the harvesting capacity is developed to examine. In general, leveraging the proposed moment matching method, the Cox constellation approximates the polar constellation. A more detailed comparison with other constellations is beyond this paper's scope and will be explored in future work.


\section{Conclusion and Future Work}
Leveraging a Cox point process model that integrates LEO satellites' spatial distributions and their motions, this paper proposes an analytical framework to investigate the space-time correlated behavior of a delay-tolerant data harvesting architecture leveraging LEO satellites. We derive various performance metrics including harvest time fraction, the harvesting data per pass, the harvesting capacity, and delay distribution.  By presenting these key performance metrics as the network's distributional parameters, this paper provides intuitions and insights on how to design, deploy, and optimize such a delay-tolerant LEO satellite network. 

Within the context of stochastic geometry, this work is a pioneering attempt to examine the spatially correlated motion of LEO satellites under a unifying framework. For future work, the analytical framework extends to better represent real LEO satellite deployment scenarios. To name a few, such examples are (i) representing a group of regularly spaced LEO satellites on each orbit as a comb point process to describe spatial repulsion of LEO satellites, (ii) varying small-scale fading w.r.t. the elevation angle of the LEO satellite harvester, (iii) evaluating and comparing the performance metrics to polar constellations, (iv) capturing a spatially consistent blockage based on real obstacles, and (v) considering the impact of inter or intra satellite interference for the the harvesting capacity evaluation.

		\appendices
		\section{Proof of Lemma \ref{lemma1}}\label{A:0}
			The typical user can communicate with any satellites at distances less than $\gamma.$  An orbit is communicable if it can have any LEO satellite at distance less than $\gamma.$ Therefore, by definition, the inclination of an effective communication orbit must be such that  $\pi/2-\xi\leq \phi \leq \pi/2+\xi $ where $\xi=\arccos\left((r_o^2+r_e^2-\gamma^2)/(2 r_or_e)\right) $. Since $\cO$ is a Poisson on rectangle $\cR, $ the number of such orbits follows  a Poisson random variable with its mean---$\lambda $ times the area of the set whose inclination is between $\pi/2-\xi, $ and $\pi/2+\xi$. Therefore, its mean is given by ${\lambda}\sin (\xi)$.
			
		On the other hand, based on Section \ref{S:2-Prelim}, the segment length of the orbit $l(\theta,\varphi)$ that meet the cap $\cS_\gamma$ is given by Eq. \eqref{arc:length}. Since satellites are distributed as a Poisson point process conditionally on each orbit, the number of the satellites on the segment $\cA(\theta,\phi)$ is given by the product of $\mu/(2\pi r_o)$ and $2 r_o \arccos(\sqrt{1-\cos^2(\xi)\csc^2(\phi)})$. Hence, the average number of satellite harvesters within $\gamma $ is 
		\begin{align}
			\bE[N] &= \bE\left[\sum_{(\theta,\phi)\in\cO}\bE\left[\left. \ind_{\sum_{X_i\in\psi_{\theta,\varphi}}} \ind_{\|X_i-(0,0,r_e)\|<\gamma} \right|\cO\right]\right]\nnb\\
			&= \bE\left[\sum_{(\theta,\phi)\in\cO}\frac{\mu}{\pi}\arcsin(\sqrt{1-\cos^2(\xi)\csc^2(\phi)})\right]\nnb.
		\end{align}
		Then, by employing Campbell's mean value theorem, we have 
		\begin{align}
			\bE[N] &=\frac{\lambda\mu}{2\pi}\int_{\pi/2-\xi}^{\pi/2+\xi}\!\!\!\! \sin(\phi)\arcsin(\sqrt{1-\cos^2(\xi)\csc^2(\phi)})\diff \phi\nnb\\
			&=\frac{\lambda\mu}{\pi}\int_{0}^{\xi} \cos(\varphi)\arcsin(\sqrt{1-\cos^2(\xi)\sec^2(\varphi)})\diff \phi\nnb
		\end{align}
		where we use the change of variables $\phi=\pi/2-\varphi$. Note that we have $\xi = \arccos\left((r_o^2+r_e^2-\gamma^2)/(2 r_or_e)\right).$ 
	\section{Proof of Theorem \ref{T:1}}\label{A:1}
		The satellite point processes conditionally on the orbit process are ergodic \cite{last2017lectures}. 
	Thanks to its ergodic and time invariant property, the fraction of time that the typical user has no satellite on the orbit $l(\theta,\phi)$ is the same as the probability that the typical user does not have any satellite within its sensing distance $\gamma$ at any given time. Using Section \ref{S:2-Prelim} and Eq. \eqref{arc:length}, the probability that there is no LEO satellite within distance $\gamma$ is 
	\begin{align}
		\exp\left(-\frac{\mu}{\pi}\arcsin\left(\sqrt{1-{\cos^2(\xi)\csc^2(\phi)}}\right)\right).
	\end{align}
	For each time, the typical user has either at least one LEO satellite in $\cS_\gamma$ or not. For a given time, the typical user has no LEO satellite if and only if there is no LEO satellite for each and every orbit. As a result, using the independence of the orbits in $\cO$, $\cT$ the fraction of time that the typical user has any LEO satellite within distance $\gamma$ is given by 
	\begin{align}
		&1-\bP\left(\prod_{(\theta,\phi)\in \cO} \exp\left(-\frac{\mu}{\pi}\arcsin(\sqrt{1-{\cos^2(\xi)\csc^2(\phi)}})\right) \right)\nnb\\
		&=1-e^{-{\lambda}\int_{\pi/2-\xi}^{\pi/2+\xi} \frac{\sin(\phi)}{2}\left(1- e^{-\mu/\pi \arcsin(\sqrt{1-\cos^2(\xi)\csc^2(\phi)})}\right)\diff \phi }\nnb\\
		&=1-e^{-{\lambda}\int_0^{\xi} \cos(\varphi)\left(1- e^{-\mu/\pi \arcsin(\sqrt{1-\cos^2(\xi)\sec^2(\varphi)})}\right)\diff \varphi },
	\end{align}
	where we use the probability generating functional of the Poisson point process of intensity $\lambda\sin(\phi)/2\pi$ defined on the rectangle set $\cR:=[0,\pi]\times[0,\pi]$. Since only orbits with angles $\pi/2-\xi<\phi<\pi/2+\xi$  meets the spherical cap $\cS_\gamma,$ the integration w.r.t. $\phi$ is from $\pi/2-\xi$ to $\pi/2+\xi$. Then we employ the change of variables to get the final result. 
	
	\section{Proof of Theorem \ref{T:2}}\label{A:2}
		First, the average successful data harvest per pass is obtained by integrating the fixed rate over the duration of time that the association satellite passes over the typical user. Let $m$ and $\tau$ be the modulation rate and the SNR threshold, respectively. Let $B_w$ be the bandwidth of the data harvesting transmission. Then, 
	for the satellite on an orbit with longitude $\theta$ and inclination $\phi$, the amount of successful data transmission per pass is 
	\begin{equation}
		mB_w (t_2-t_1) \bP(\SNR_{t_1\to t_2}>\tau|\theta,\phi) \label{16}, 
	\end{equation} 
	where $t_1 $ is the time that the typical user begins to upload and $t_2 $ is the time that the typical user ceases to upload. Here, $\omega_1$ and $\omega_2$ are two arguments corresponding to those instances, respectively. Then, we get $ (t_2-t_2) = (\omega_2-\omega_1)/\omega_s$ where $\omega_s$ is the angular speed of the LEO satellite. In Eq. \eqref{16}, $\SNR_{t_1\to t_2}$ means the SNR from time $t_1$ and $t_2$ and hence $\bP(\SNR_{t_1\to t_2}>\tau|\theta,\phi)$ is given by a function of $\omega $. Therefore, by using the fact that the locations of satellites is modeled as a Poisson point process on each orbit, we simply get 
	\begin{align}
		&\bP(\SNR_{t_1\to t_2} > \tau|\theta,\phi) \nnb\\
		&= \bE_\omega \left[\bP(\SNR_{\omega}>\tau|\omega,\theta,\phi)\right]\nnb\\
		&= \frac{1}{\omega_2-\omega_1}\int_{\omega_1}^{\omega_2}\bP(\SNR_{\omega}>\tau|\theta,\phi,\omega)\diff \omega\nnb\\
		&=\frac{\int_{\omega_1}^{\omega_2} \overbar{F_H}\left({\frac{\tau \sigma^2}{pg} (r_o^2-2r_or_e\sin(\omega)\sin(\phi)+r_e^2)^{\frac{\alpha}{2}}}\right)\diff \omega}{\omega_2-\omega_1},\label{17}
	\end{align}
	where $\SNR_{\omega}$ is the SNR of the satellite having the argument angle $\omega$, $\sigma^2 $ is the noise power, $g$ is the aggregate antenna gain, and $\overbar{F_H}(x) $ is the CCDF of the $H$ evaluated at $x.$ Simply, $pg/\sigma^2$ is the received SNR at distance $1$ meter. 
	
	To get Eq. \eqref{17}, we use Section \ref{S:2-Prelim} to obtain the argument angles $\omega_1= \pi/2 - \arcsin(\sqrt{1-\cos^2(\xi)\csc^2(\phi)}) $ and $\omega_2= \pi/2 + \arcsin(\sqrt{1-\cos^2(\xi)\csc^2(\phi)})$. 
	
	Combining Eqs. \eqref{16} and \eqref{17}, we have 
	\begin{align}
		\frac{mB_w}{\omega_s}\int_{0}^{\omega_0} \!\! \overbar{F_H}\left(\frac{\tau \sigma^2}{pg} (r_o^2-2r_or_e\cos(\omega)\sin(\phi)+r_e^2)^{\frac{\alpha}{2}}\right)\diff \omega ,\label{255}
	\end{align}
	where we use the change of variables to get $\omega_0 =\arcsin(\sqrt{1-\cos^2(\xi)\csc^2(\phi)})  $.  Note Eq. \eqref{255} is a function of $\phi $ and $\phi \in(\pi/2-\xi,\pi/2+\xi)$, where $\pi/2+ \xi$ is the maximum inclination of the orbits intersecting $\cS_\gamma$. The unit of Eq. \eqref{255} is bits.
	
	\par With an adaptive modulation and coding, the rate is $\log(1+\SNR)$, where $\SNR$ is the instantaneous SNR. For orbit $l(\theta,\phi),$ the harvest data  per pass is 
	\begin{align}
		B_w (t_2-t_1)\bE\left[\log_2\left(1+\SNR\right)|\theta,\varphi\right].\nnb \label{19}
	\end{align}	
	Since $\log(1+\SNR) $ is a positive random variable, the above expectation is given by 
	\begin{align*}
		&\bE\left[\log_2\left(1+\SNR\right)|\theta,\phi\right]\\
		& =\int_0^\infty \bP(\log_2(1+\SNR)>r|\theta,\phi) \diff r \\
		&= \int_{0}^{\infty} \bP(\SNR>2^r-1|\theta,\phi) \diff r \\
		&= \int_{0}^{\infty} \int_{\omega_1}^{\omega_2}\left[\bP(\SNR_{\omega}>2^r-1|\theta,\phi,\omega)\right] \diff \omega\diff r,
	\end{align*}
	where $\SNR_\omega$ is the SNR of the satellite harvester when the satellite's argument angle is $\omega$. 

	Now using Eqs. \eqref{17} and \eqref{255}, with an adaptive modulation and coding, the average amount of data transmission per pass from the satellite on orbit $l(\theta,\phi)$ is given by 
	\begin{align}
		&\frac{B_w}{\omega_s}\int_{0}^{\infty}\int_{0}^{\omega_0}  \overbar{F_H}\left(\frac{\sigma^2(2^r-1)}{pg}\right.\nnb\\
			&\hspace{20mm}\left. (r_o^2-2r_or_e\cos(\omega)\sin(\phi)+r_e^2)^{\frac{\alpha}{2}}\right)\diff \omega \diff r.
	\end{align}
	
	\section{Proof of Theorem \ref{T:3}}\label{A:3}
		By definition, the harvesting capacity is given by the achievable rate from typical user to its association LEO satellite harvester. At any given time, we have 
	\begin{align}
		\bar{r} &= \bE\left[\log_2(1+\SNR)\right]
		= \int_0^\infty \bP(\SNR>2^v-1) \diff v\label{eq:23}
	\end{align}
	where we use the fact that the SNR is a positive random variable. Since the typical user is associated with its closest LEO satellite, the coverage probability is 
	\begin{align}
		&\bP(\SNR>\tau) \nnb\\
		& = \bE\left[\ind_{\SNR>\tau}\right]\nnb\\
		&=\bE_{\cO}\left[\bE_{l_\star}\left[\bE\left[\left.\ind_{\SNR>\tau}\right|  l_\star, \cO \right]\right]\right]\nnb\\
		&=\bE_{\cO}\left[\bE_{l_\star}\left[\int_{}^{}\bP\left(\left.{{H}>\frac{\tau u^\alpha}{pg/\sigma^2}} \right| r, l_\star,\cO\right) f_r(u)\diff u\right]\right]\nnb\\
		&=\bE_{\cO}\left[\bE_{l_\star}\left[\int_{r_o-r_e}^{\gamma}\overbar{F_H}\left(\frac{\tau\sigma^2u^\alpha}{pg}\right) f_r(u)\diff u\right]\right]\label{25},
	\end{align}
	where we condition on the orbit process $\cO$ and then on the orbit $l_\star$. Here, $l_\star = l(\theta_\star,\phi_\star)$ is the orbit that has the satellite harvester closest to the typical user.  $\overbar{F_H}(x)$ is the CCDF of the random variable $H$ evaluated at $x$ and $f_r(u)$ is the probability density function of  the distance to the closest LEO satellite, conditional on $\cO$ and $l_\star$. 
	
	To get the distribution function $f_r(u)$, for $r_a<u<\gamma $, with $r_a$ the satellite altitude, we write 
	\begin{align}
		f_r(u) & = \frac{\diff }{\diff r}\bP(\|X_\star\| \leq u|\cO,l_\star)\nnb\\
		&=\bP(\Psi(B_{0,0,r_e}(u))=\emptyset|\cO) \nnb\\ 
		& \hspace{5mm}\frac{\diff }{\diff r}\left(1- \bP(\|X_\star\| > u|X_\star\in\psi_\star)\right)\nnb\\
		&=\frac{\diff }{\diff r}\left(1- g(\varphi_\star,u)\right)\prod_{\phi_{i}\in\cO}^{\phi<\kappa}g(\phi_{i},u),
	\end{align}
	where we use that the association LEO satellite harvester is on the orbit $l_\star. $ For simplicity,  
	\begin{align}
		\kappa &= \arccos\left(\frac{r_o^2+r_e^2-u^2}{2r_or_e }\right),\\
		g(\phi, u )  &= e^{{\left.-\frac{\mu}{\pi}\arcsin\left(\sqrt{1-\cos^2(\kappa)\csc^2(\phi)}\right)\right.}},\\
		\bP(1-g(\phi_\star,u))' &= \frac{\mu u \csc(\phi_\star)g(\phi_\star,u)}{\pi r_e r_o \sqrt{1-\cos^2(\kappa)\csc^2(\phi_\star)}}.\label{29}
	\end{align}
	We combine Eqs. \eqref{25}--\eqref{30} and then use Fubini's theorem to arrive at 
	\begin{align}
		&\bP(\SNR>\tau )\nnb\\
		& = \int_{r_a}^\gamma \frac{\mu u \overbar{F_H}\left(\!\frac{\tau \sigma^2u^\alpha}{pg}\!\right)}{\pi r_e r_o} \!\bE_{\cO,l_\star}\!\left[\frac{ \csc(\phi_\star)\prod\limits_{\phi_{i}\in\cO+\phi_\star}^{\phi<\kappa}g(\phi_{i},u)}{ \sqrt{1-\cos^2(\kappa)\csc^2(\phi_\star)}}\right]\diff u.\label{31}
	\end{align}
	In the product form of orbits, we have ${\phi_{i}\in\cO+\phi_\star}$ that is the independent superposition of $\cO$ all orbits and $l_\star$ the orbit containing the association LEO satellite. Then, since $\cO$ and $l_\star$ are independent processes, we use the Campbell's averaging formula and the probability generating functional of Poisson point process to get the following integral formulas. 
	\begin{align}
		&\bE_{l_\star}\left[\frac{|\csc(\phi_\star)|g(\phi_\star,u)}{\sqrt{1-\cos^2(\kappa)\csc^2(\phi_\star)}}\right]\nnb\\
		&\hspace{5mm}=\frac{\lambda}{2}\int_{\pi/2-\kappa}^{\pi/2+\kappa} \sin(x)\frac{\csc(x)g(x,u)}{\sqrt{1-\cos^2(\kappa)\csc^2(x)}} \diff x \nnb\\
		&\hspace{5mm}=\lambda\int_{0}^{\kappa} \frac{\bar{g}(x,u)}{\sqrt{1-\cos^2(\kappa)\sec^2(x)}} \diff x ,\label{eq:31}
	\end{align}
	where we use the change of variables and let $\bar{g(x,u)} $ be as follows: 
	\begin{equation}
	\bar{g}(x,u ) = e^{\left.-\frac{\mu}{\pi}\arcsin\left(\sqrt{1-\cos^2(\kappa)\sec^2(x)}\right)\right.}.\label{30}
\end{equation}
Moreover, by employing the probability generating functional of the orbit process, we have 
	\begin{align}
		&\bE_{\cO}\left[\prod_{\phi_{i}\in\cO}^{|\phi-\pi/2|<\kappa}g(\phi_{i},u)\right] \nnb\\
		&= \exp\left(-\frac{\lambda}{2}\int_{\pi/2-\kappa}^{\pi/2+\kappa} \sin(\phi) \left(1-{g}(\phi,u) \right)\diff \phi\right)\nnb\\
		&= \exp\left(-{\lambda}\int_{0}^{\kappa} \cos(\varphi)\left(1-\bar{g}(\varphi,u)\right) \diff \varphi\right).\label{eq:32}
	\end{align}

\par As a result, by substituting Eqs. \eqref{eq:31} and \eqref{eq:32} into \eqref{31}, the coverage probability with threshold $\tau$ is given by 
	\begin{align}
		&\bP(\SNR>\tau)\nnb\\
		&=\int_{r_o-r_e}^{\gamma}\frac{\lambda\mu u \overbar{F_H}\left(\frac{\tau u^\alpha}{pg/\sigma^2}\right) \left.\int_0^\kappa \frac{\sec(x)\bar{g}(x,u)}{\sqrt{1-\cos^2(\kappa)\sec^2(x)}} \diff x\right.}{\pi r_e r_o \exp\left({{\lambda}\int_0^\kappa \cos(\varphi)\left(1-\bar{g}(\varphi,u)\right)\diff \varphi}\right)}\diff u.
	\end{align}
	Finally, we use Eq. \eqref{eq:23} to derive the harvesting capacity of the proposed network architecture leveraging the above coverage probability expression. 
	
	\section{Proof of Theorem \ref{T:4}}\label{A:4}
		To obtain the CCDF of delay random variable $D,$ we use the fact that $\bP(D>d)$ is the probability that there is no satellite harvester at distance less than $\gamma$ within the time duration $d.$ We have 
	\begin{align}
		&\bP(D>d) \nnb\\
		&= \bP(\text{no satellite at distance $\gamma$ over time $d$} )\nnb\\
		&= \bP\left(\prod_{{(\theta,\phi)}\in\cO}\bP\left(\text{no satellite of orbit $l(\theta,\phi)$}\right.\right.\nnb\\
		&\hspace{28mm}\left.\left.\text{at distance $\gamma$ over time $d$} |\cO\right)\right)\label{eq:delaybasis}.
	\end{align}
	\par For an orbit $l(\theta,\phi)$, this event corresponds to the event that there is no point over the time $d$ on the arc $\cA(\theta,\phi) $. Then, since satellites move at the angular speed of $\omega_s$ on the orbit, the probability is the chance that at time zero, there is no LEO satellite on the segment $\cA(\theta,\varphi) \cup \bar{l}_d (\theta,\varphi)$, where $\bar{l}_{d}(\theta,\varphi)$  is the fraction of the orbit that corresponds to the travel distance of satellite over the time $d.$
	
	Since the travel distance over the time $d$ is $r_s\omega_sd$ and the length of $\cA(\theta,\phi)$ is given by Eq. \eqref{arc:length}, the conditional probability of Eq. \eqref{eq:delaybasis} is given by 
	\begin{equation}
		\exp\left(-\frac{\mu}{2\pi}\left(2\arcsin\left(\sqrt{1-\cos^2(\xi)\csc^2(\phi)}\right)+\omega_s d \right)\right),\label{m1}
	\end{equation}
	where we use the void probability of the Poisson point process of intensity $\mu/(2\pi r_o)$. 
	
	
	Therefore, 
	the delay distribution of the proposed network is given by 
	\begin{align}
		&\bP(D>d)\nnb\\ 
		&=\bE\left[\prod_{ \cO}^{\phi<\xi}e^{-\frac{\mu}{2\pi}\left(\arcsin\left(\sqrt{1-\cos^2(\xi)\csc^2(\phi)}\right)+\omega_s d \right)}\right]\nnb\\
		&=e^{-{\lambda}\int_{\pi/2-\xi}^{\pi/2+\xi} \frac{\sin(\phi)}{2}\left(1 - e^{-\frac{\mu}{2\pi}\left(2\arcsin\left(\sqrt{1-\cos^2(\xi)\csc^2(\phi)}\right)+\omega_s d \right)}\right)\diff \phi},\nnb\\
		&=e^{-{\lambda}\int_{0}^{\xi} {\cos(\phi)}\left(1 - e^{-\frac{\mu}{2\pi}\left(2\arcsin\left(\sqrt{1-\cos^2(\xi)\sec^2(\varphi)}\right)+\omega_s d \right)}\right)\diff \varphi},\nnb
	\end{align}
	where the integration w.r.t. $\phi$ is from $\pi/2-\xi$ to $\pi/2+\xi$ where $\pi/2-\xi$ is the minimum inclination of the orbit that meets the spherical cap $\cS_\gamma$ and $\pi/2+\xi$ is the maximum inclination of the orbit that meets the spherical cap $\cS_\gamma$. Then, by applying the probability generating functional of the orbit process on the rectangle $\cR$ and by employing the change of variable, we obtain the final result.

\section*{Acknowledgment}
This work was supported by the National Research Foundation of Korea grant funded by the Korea government. (No. 2021R1F1A1059666).

\bibliographystyle{IEEEtran}
\bibliography{ref}

\end{document}